\documentclass[
twocolumn,
groupedaddress,
 amsmath,amssymb,
 aps,
prl,
]{revtex4-2}

\usepackage{graphicx}
\usepackage{dcolumn}
\usepackage{bm}
\usepackage{hyperref}
\usepackage[separate-uncertainty=true]{siunitx}
\usepackage{subcaption}
\DeclareCaptionLabelFormat{bf-parens}{(\textbf{#2})}
\usepackage{comment}
\usepackage{placeins}
\usepackage{titletoc}

\titlecontents{section}[0em]{}{\bfseries\thecontentslabel\quad}{}{\titlerule*[0.5pc]{.}\contentspage}
\titlecontents{subsection}[2em]{}{\thecontentslabel\quad}{}{\titlerule*[0.5pc]{.}\contentspage}

\DeclareSIUnit\molar{\mole\per\cubic\deci\metre}
\DeclareSIUnit\Molar{\textsc{m}}

\begin{document}


\title{Breakdown of the optical saturation regime in  molecular single-photon emitters}

\author{Hugo Levy-Falk$^1$}
\email{hugoloup.levy-falk@ino.cnr.it}
\author{Daniele De Bernardis$^{1,2}$}
\author{Elena Fanella$^{3}$}
\author{Louise Morla\"{e}s$^{1,2}$}
\author{Costanza Toninelli$^{1,2}$}
\email{toninelli@lens.unifi.it}

\affiliation{National Institute of Optics (CNR-INO), c/o LENS via Nello Carrara 1, Sesto F.no 50019, Italy}
\affiliation{European Laboratory for Non-Linear Spectroscopy (LENS), Via Nello Carrara 1, Sesto F.no 50019, Italy}
\affiliation{Physics Department, University of Naples, Via Cinthia 21, Fuorigrotta 80126, Italy}


\date{\today}

\begin{abstract}
Solid-state single organic molecules, such as dibenzoterrylene (DBT) in organic matrices, are prominent deterministic single-photon sources, usually modeled as effective two-level systems (TLS). We show that single DBT molecules in anthracene nanocrystals, under strong continuous-wave driving, depart from this picture: instead of the expected saturation, fluorescence is strongly suppressed at high resonant pump power, while the linewidth broadens beyond the TLS prediction -- an \emph{anomalous saturation} regime. Comparing coherent and incoherent excitation and independently calibrating temperature via phonon-induced dephasing rules out laser-induced heating and intersystem crossing. Instead, a model of intensity-dependent excited-state absorption (ESA) toward a short-lived dark state quantitatively reproduces both the fluorescence suppression and the linewidth broadening. We further show that matrix quality mitigates this quenching, and that pulsed excitation schemes are strategic toward full population inversion, with direct implications for quantum nanophotonics and molecular optomechanics.
\end{abstract}

\maketitle


\message{Column width \the\columnwidth}

\message{Line width \the\linewidth}

\message{Text width \the\textwidth}


The interaction of light with isolated quantum systems lies at the core of quantum science and technologies, from the foundations of quantum optics~\cite{cohentannoudji_atomphoton_1998} to modern applications in quantum communication, sensing, and information processing~\cite{Labonte_PRXQuantum.5.010101,Degen_RevModPhys.89.035002, Lavroff2021}. 
Particularly attractive are solid-state quantum emitters, offering strong optical transitions with the opportunities of device integration and potential scalability. 
A broad range of platforms has emerged over the last few decades, including semiconductor quantum dots, color centers in wide band-gap semiconductors, rare-earth ions and molecular emitters~\cite{PelayoScience2021, Rodgers2021, Zhong2017, Bayliss2020, toninelliSingleOrganicMolecules2021, Heindel:23}. Among these, organic molecules in solid matrices combine high fluorescence quantum yield with long coherence times of their zero-phonon line (ZPL) at cryogenic temperature~\cite{toninelliSingleOrganicMolecules2021}, making them attractive deterministic single-photon sources~\cite{Shkarin2026}. Dibenzoterrylene (DBT) molecules embedded in anthracene (Ac) crystals in particular exhibit high photon indistinguishability~\cite{trebbia_indistinguishable_2010, lombardi_triggered_2021, duquennoy_real-time_2022}, efficient integration with cavities and waveguides~\cite{Wang2019, Lange2026, Huang2025, Boissier2021, Fournier2026}, and additional handles such as optical frequency tuning~\cite{colauttiLaserInducedFrequencyTuning2020} and confinement in submicron crystals~\cite{pazzagli_self-assembled_2018, Musavinezhad2024}.

These emitters are almost universally described as effective two-level systems (TLS)~\cite{Wang2019}, absorbing and emitting one photon at a time, with an optical response that saturates at high continuous-wave (CW) driving intensity. Beyond saturation, nonlinear quantum-optical phenomena become relevant, such as the Mollow triplet~\cite{Mollow_PhysRev.188.1969,Kimble_PhysRevA.13.2123,Wrigge2008}. In this description, the molecule's residual decay channels besides the ZPL, such as non-radiative ones or Stokes shifted emission, are captured by the effective decay rate of a damped two-level system. With the excitation of molecular vibrations, strong driving further enables transistor-like optical behavior~\cite{Hwang2009}, four-wave mixing~\cite{Maser2016}, and coherent optomechanical interactions between electronic and vibrational degrees of freedom~\cite{zirkelbach_spectral_2023, De_Bernardis2025, roelli_molecular_2020, Chen2021, Xomalis2021}, with applications extending to nanoscale sensing~\cite{estesoQuantumThermometrySingle2023, moradi_matrixinduced_2019, Ciancico2025}. Still, the optical saturation of the fluorescence rate remains a valid prediction also in the configuration of vibration-assisted pumping. However, precisely because of this rich internal structure, operating molecular emitters in the high-saturation regime may stimulate unpredicted photophysical paths beyond the simple two-level picture -- a question of both fundamental and applied relevance that remains largely unexplored.

In this Letter, we experimentally investigate single DBT molecules in anthracene nanocrystals (NCs) in the high-saturation regime and provide strong evidence for the failure of the two-level description, even under resonant ZPL excitation. We identify an intrinsic feature that we refer to as the \emph{anomalous saturation}: instead of approaching the standard two-level saturation limit, DBT fluorescence is strongly suppressed at high excitation intensities, while the optical linewidth broadens beyond the TLS prediction. We show that this behavior cannot be explained by laser-induced heating of the host matrix nor by a power-dependent intersystem-crossing yield. Instead, we introduce a model based on an intensity-dependent depletion channel of the excited state, compatible with excited-state absorption (ESA), which quantitatively accounts for the observed fluorescence suppression and linewidth broadening. We further discuss how this channel limits the achievable excited-state population under CW driving, and how matrix optimization and pulsed excitation schemes can mitigate the effect. These findings unveil a previously unexplored optical regime for molecular single-photon emitters under strong CW excitation, with implications for the validity of two-level descriptions and operation guidelines in quantum optics and nanophotonics application. Extended derivations, methods, and supporting data are provided in the Supplemental Material~\cite{SupplementalMaterial}.

{\it Experimental Results.}---
\begin{figure*}
    \centering
    \begin{subcaptiongroup}
        \subcaptionlistentry{resonant saturation scan}
        \label{fig:resonant saturation scan}
        \subcaptionlistentry{resonant saturation amplitude}
        \label{fig:resonant saturation  amplitude}
        \subcaptionlistentry{resonant saturation broadening}
        \label{fig:resonant saturation broadening}
        \subcaptionlistentry{cartoon anomalous saturation}
        \label{fig:cartoon anomalous saturation}
    \end{subcaptiongroup}
    \includegraphics[width=\linewidth]{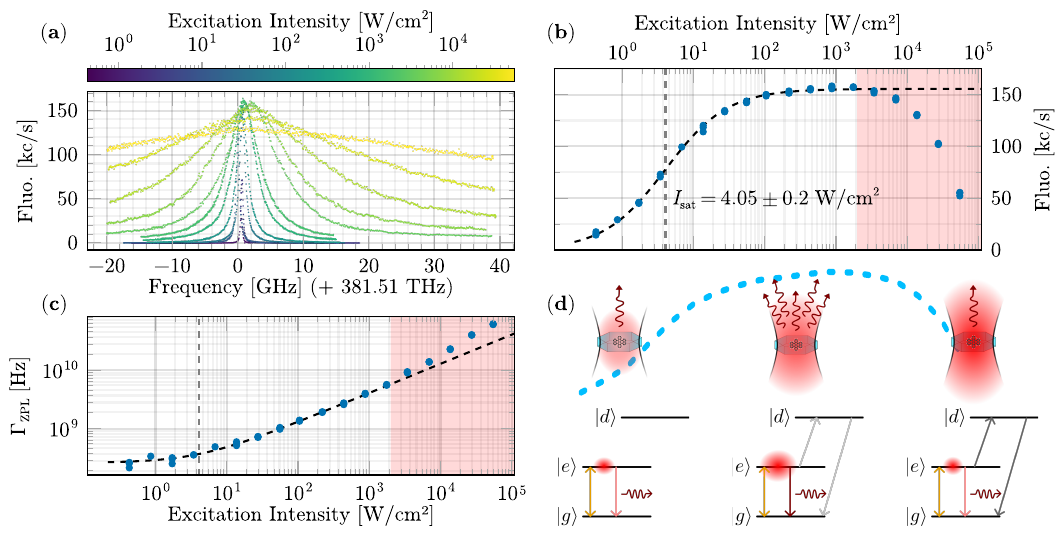}
    \caption{\subref{fig:resonant saturation scan}~Resonant scans of the ZPL performed at various laser excitation intensities while collecting red-shifted photons. \subref{fig:resonant saturation  amplitude}~Amplitudes extracted from the fit with a Lorentzian line on the resonant scans (blue dots). The anomalous regime is highlighted in light red. Result of a fit procedure with a standard saturation law (Eq.~\ref{eq:standard saturation}), yielding a saturation intensity of \SI{4.05\pm0.2}{W/cm^2} (black dashed line).  \subref{fig:resonant saturation broadening}~FWHM of the fitted Lorentzian lines (blue dots); the dashed black line represents the fit results with a standard TLS model (see main text). \subref{fig:cartoon anomalous saturation}~Cartoons describing the anomalous saturation pattern for DBT:anthracene nanocrystals: as the excitation intensity increases, the excited-state ($|e\rangle$) population grows resulting in a higher emission rate, until it saturates. For stronger drive though, fluorescence is quenched. We suggest this is due to absorption from the excited state into a highly dissipative dark state ($|d\rangle$).}
    \label{fig:2}
\end{figure*}
All measurements are performed on a home-built epifluorescence microscope (see Supplemental Material (SM)~\cite{SupplementalMaterial}, Sec.~\ref{sec:methods}, for setup and sample details).



{\it 0-0 Excitation.}---
Single molecules are excited coherently via their HOMO-LUMO ${\rm S}_{0-0}$ transition (ZPL at $\sim\SI{785}{nm}$), and the Stokes-shifted fluorescence is collected on a single-photon avalanche photodiode, as the laser frequency is scanned across the \SI{50}{MHz}-wide resonance of an isolated molecule (Fig.~\ref{fig:resonant saturation scan}). At low pump power the lineshape is well described by a Lorentzian profile~\cite{loudonQuantumTheoryLight2000}. From Lorentzian fits at each intensity (SM~\cite{SupplementalMaterial}, Fig.~\ref{fig:sup:resonant scans anomalous saturation}), we extract the fluorescence amplitude and full width at half maximum (FWHM), shown as blue dots in Fig.~\ref{fig:resonant saturation  amplitude} and Fig.~\ref{fig:resonant saturation broadening}. 
The standard TLS saturation law reads
\begin{equation}
    R^{\rm res}(I) = R^{\rm res}_\infty \frac{I}{I + I_\text{sat}^{\rm res}}, \label{eq:standard saturation}
\end{equation}
with $R$ the emission rate, $I$ the laser intensity, $I_\text{sat}^{\rm res}$ the saturation intensity, and $R_\infty^{\rm res}$ the maximum fluorescence rate. This law fits the data up to $\sim\SI{e3}{W/cm^2}$ (black dashed line, fit restricted to $I<\SI{2e3}{W/cm^2}$), but a clear deviation and unexpected drop are
observed in the high saturation regime. 

The ZPL linewidth is also compared to the optical Bloch equation prediction (dashed line, Fig.~\ref{fig:resonant saturation broadening}; see SM~\cite{SupplementalMaterial}, Sec.~\ref{sec:OBE_res_pump}), giving the power-broadened FWHM
$\Gamma_{\rm zpl} = \frac{\gamma_2}{\pi}\sqrt{1 + S^{\rm res}},$
where $S^{\rm res}=I/I_{\rm sat}^{\rm res}$ is the saturation parameter and $2\gamma_2=\gamma_1+\gamma_\phi$ is the coherence decay rate (inverse lifetime $\gamma_1$ plus dephasing rate $\gamma_\phi$). Again, a clear deviation sets in above $\sim\SI{20}{kW/cm^2}$. We refer to this departure from the standard TLS description as the \emph{anomalous saturation regime}. Strikingly, after reaching the expected saturation plateau, the molecule becomes dimmer as the excitation intensity is increased further, instead of remaining constant. This points to a competing, intensity-dependent loss channel that sets in only at high intensity and progressively depletes the excited state, as we investigate below (Fig.~\ref{fig:cartoon anomalous saturation}).


{\it 0-1 Excitation.}---
To probe this effect independently, we pump the molecule incoherently via its anti-Stokes ${\rm S}_{0-1}$ transition at \SI{767}{nm} and detect fluorescence in a \SI{0.4}{nm}-narrow window around the ZPL (Fig.~\ref{fig:off-resonant saturation}, inset). The experimental data (blue dots in Fig.~\ref{fig:off-resonant saturation}) show that in this case the fluorescence never reaches full saturation and rapidly decays for high values of the excitation laser power, hence a fit with standard TLS is not obvious. This behavior is highly reproducible: it is observed in more than 20 molecules across different NCs and, after rescaling for individual alignment, saturation intensity, and quantum yield, all curves collapse onto a common trend, ruling out artefacts (SM~\cite{SupplementalMaterial}, Sec.~\ref{sec:uni_offres_AS}). We further confirm that the fluorescence drop is not a spectral redistribution (consistent reduction is seen integrating different spectral windows) and, most importantly, that it is completely reversible with laser power, excluding photobleaching (SM~\cite{SupplementalMaterial}, Secs.~\ref{sec:stokes_shifted} and \ref{sec:reversibility}). Comparing the two schemes, the saturation parameter $S=R(I)/(R_\infty-R(I))$ reaches $S^{\rm res}\approx100$ under resonant driving -- consistent with the excited-state population approaching its TLS maximum $\rho_{ee}\approx0.5$ -- whereas $S^{\rm off}$ is bound to smaller values $\in[0.1,1]$ under off-resonant pumping, i.e. $\rho_{ee}\lesssim0.5$ despite a higher absolute count rate (see SM~\cite{SupplementalMaterial}, Sec.~\ref{sec:S_param_appendix} for the full analysis and extracted populations). In the next sections, we discuss the origin of the \emph{anomalous saturation} effect.
\begin{figure}
    \centering
    \includegraphics[width=\linewidth]{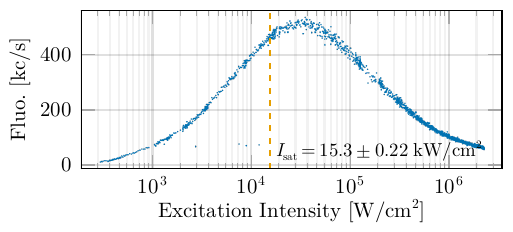}
    \caption{Anomalous saturation curve (blue dots) in off-resonant configuration (inset).}
    \label{fig:off-resonant saturation}
\end{figure}

{\it Ruling out temperature.}---
A natural hypothesis is that the anomalous saturation results from laser-induced heating: although the Ac matrix is transparent at \SI{785}{nm}, the gold substrate could locally warm up beyond the cryostat's cooling capability, exposing the molecule to an effective, intensity-dependent temperature $T_{\rm eff}(I)$. Expanding the theory presented in  Ref.~\cite{clear_phonon-induced_2020}, we model the resulting phonon-induced dephasing, Debye-Waller (DW) suppression of the Rabi frequency, and thermal frequency shift (SM~\cite{SupplementalMaterial}, Sec.~\ref{sec:thermal_theory}), yielding a temperature-dependent saturation parameter
\begin{equation}\label{eq:S_param_thermal}
    S^{\rm res}(I,T) = \frac{I}{I_{\rm sat}^{\rm res}} \frac{e^{-2F(T) }}{1+\gamma_{\phi}(T)/\gamma_1},
\end{equation}
with $F(T)$ being the Huang-Rhys factor and $\gamma_\phi(T)$ the phonon-induced dephasing rate. Since $\gamma_\phi(T)$ directly calibrates temperature via the ZPL linewidth~\cite{estesoQuantumThermometrySingle2023}, we extract $T_{\rm eff}(I)$ from the anomalous saturation scan of Fig.~\ref{fig:2} and compare the extracted  DW factor from Eq.\eqref{eq:S_param_thermal} with the values obtained directly varying the cryostat temperature (SM~\cite{SupplementalMaterial}, Sec.~\ref{sec:heating_ruled_out}). 
\begin{figure}
    \centering
       \includegraphics[width=\linewidth]{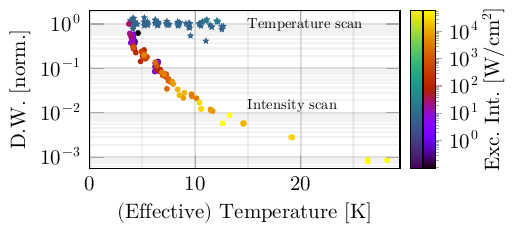}
  \caption{~Comparison of the Debye-Waller factor as a function of the cryostat temperature (blue stars) and as a function of the effective temperature of a molecule experiencing an anomalous saturation. }
    \label{fig:5f}
\end{figure}
The two data set show an orders-of-magnitude discrepancy in their temperature scaling (Fig.~\ref{fig:5f}). While genuine heating data are consistent with literature trends~\cite{clear_phonon-induced_2020},  the anomalous saturation data appear to deviate. This discrepancy is reinforced by the off-resonant experiment, where thermal dephasing does not enter the saturation parameter at all, which then depends on temperature solely through the DW factor (SM~\cite{SupplementalMaterial}, Sec.~\ref{sec:OBE_off_res}). The pronounced off-resonant fluorescence depletion can therefore hardly be attributed to heating. We conclude that a simple laser-induced heating hypothesis can be ruled out.

{\it Excited state absorption.}---
All evidence points to the existence of a power-dependent depletion channel of the excited state. We therefore introduce a short-lived dark state (or band of dark states), reached from the excited state via excited-state absorption (ESA), giving the excited state an intensity-dependent decay rate $\tilde{\gamma}_1(I) = \gamma_1 + \gamma_{\rm ESA}(I)$ (SM~\cite{SupplementalMaterial}, Sec.~\ref{sec:theory_ESA}). Assuming single-photon ESA, $\gamma_{\rm ESA}(I)= \gamma_{1}I/I_{\rm ESA}$, where $I_{\rm ESA}$ depends on the dark-state transition matrix element and dissipation rates. This dark state affects both the resonant and off-resonant schemes identically, and the anomalous saturation curves of Figs.~\ref{fig:2}-\ref{fig:off-resonant saturation} are described by the same universal expression as a function of the respective saturation parameter $S$,
\begin{equation}\label{eq:anomalous_saturation}
    \tilde{R} = R_{\infty}\frac{S}{\left(1+\xi S\right)^2 + S},
\end{equation}
with a single dimensionless \emph{anomaly parameter} $\xi$, quantifying the ESA rate at saturation relative to the standard relaxation rate. As shown in Fig.~\ref{fig:6}, this single-photon ESA model simultaneously reproduces the fluorescence, linewidth, and DW factor dependence with laser intensity. A two-photon ESA model (quadratic scaling, dashed line in Fig.~\ref{fig:anomalous broadening model}) fits markedly worse. Fitting multiple resonant datasets gives $\xi^{\rm res}\in [10^{-3}; 10^{-2}]$, while the same model applied to the off-resonant data (adjusting $R_\infty$ and re-fitting $\xi$) gives a much larger $\xi^{\rm off}\in [0.1;1]$.

This asymmetry is consistent with the model: as derived in SM~\cite{SupplementalMaterial}, Sec.~\ref{sec:theory_ESA}, where $\xi^{\rm res} = I_{\rm sat}^{\rm res}/I_{\rm ESA}$ while $\xi^{\rm off} \approx (\gamma_1/\gamma_{\rm v}) I_{\rm sat}^{\rm off}/I_{\rm ESA}$, so their ratio depends only on the $0-1$ transition Franck-Condon factor. Its small value fluctuates and naturally explains the larger off-resonant anomaly parameter. The onset of anomalous saturation, $\xi S=1$, defines an intensity threshold $I_{\rm thr}=I_{\rm sat}/\xi$. In the resonant case this occurs when the ESA contribution to the linewidth equals the inverse lifetime, and in the off-resonant case when it equals the inverse vibrational lifetime. Remarkably, this single framework quantitatively accounts for the fluorescence suppression, linewidth broadening, and the different resonant/off-resonant schemes.

What is the microscopic origin of this dark state? First-principles calculations of the DBT excited-state absorption spectrum identify a transition line near the ZPL frequency~\cite{sadeq_one-_2018}, a natural candidate for the ESA pathway. A quantitative comparison with our data is not straightforward, however, since the surrounding matrix plays a decisive role. In fact, we find that $\xi$ varies with crystal quality, and is systematically smaller for molecules embedded in high-quality sublimated crystals (SM~\cite{SupplementalMaterial}, Sec.~\ref{sec:optical_shift}). This points to a depletion channel that is not purely molecular, but shaped by the local matrix environment. Consistent with this picture, the anomaly parameter correlates with the well-known laser-induced optical frequency shift of the DBT ZPL, an effect attributed to photo-induced charge trapping in the anthracene host~\cite{colauttiLaserInducedFrequencyTuning2020,Duquennoy2024,lange_superradiant_2024, Lange2026}. This suggests that charge migration following ESA, rather than a purely intramolecular transition, may set the strength of the depletion channel.

By contrast, intersystem crossing (ISC) to the triplet manifold can be excluded as the origin of the dark state. Photon-statistics measurements show that the ISC probability \emph{decreases} with increasing excitation power, while the triplet lifetime itself stays essentially constant (SM~\cite{SupplementalMaterial}, Sec.~\ref{sec:ISC_reduction}). This trend confirms that the anomalous saturation is not simply a manifestation of the residual triplet blinking of organic emitters.

\begin{figure}
    \centering
    \begin{subcaptiongroup}
        \subcaptionlistentry{anomalous saturation model}
        \label{fig:anomalous saturation model}
        \subcaptionlistentry{amonalous broadening model}
        \label{fig:anomalous broadening model}
        \subcaptionlistentry{model debye waller factor}
        \label{fig:model debye waller factor}
    \end{subcaptiongroup}
    \includegraphics[width=\columnwidth]{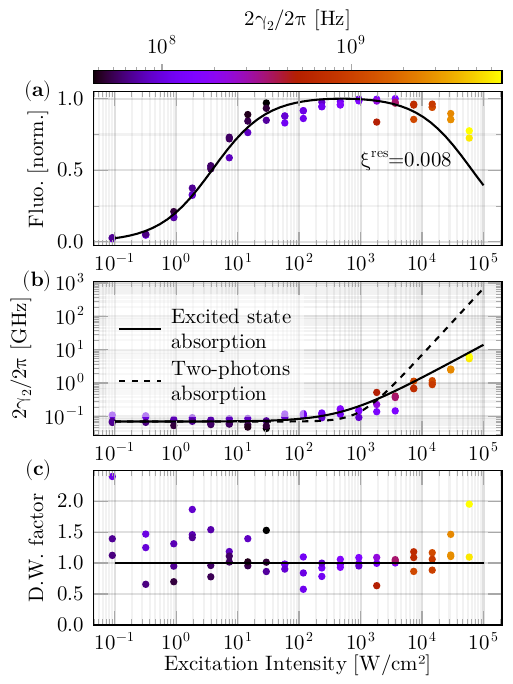}
    \caption{\subref{fig:anomalous saturation model}~Experimental saturation curve (colored dots) fitted by the model's saturation curve (solid black line). \subref{fig:anomalous broadening model}~Experimental decoherence rate of a DBT molecule as a function of excitation intensity (colored dots). The solid black line shows the behavior predicted by the excited-state absorption model. The black dashed line shows the expected behavior for a two-photons absorption model. \subref{fig:model debye waller factor}~Experimental Debye-Waller factor of a DBT molecule as a function of excitation intensity. The solid black line shows the behavior predicted by the excited-state absorption model, with $\xi^\text{res}=0.008$.}
    \label{fig:6}
\end{figure}

{\it Conclusions.}---
We report an anomalous decrease in the fluorescence of single DBT molecules in anthracene nanocrystals at high excitation intensities, observed under both coherent and incoherent pumping. Laser-induced heating can be definitively ruled out via an independent temperature calibration, while a simple model of single-photon excited-state absorption to a short-lived dark state quantitatively reproduces the fluorescence suppression and linewidth broadening in both schemes. The effect is matrix-dependent, reduced in high-quality sublimated crystals, and correlates with laser-induced optical frequency shifts, pointing to a depletion channel influenced by long-lived charge states in the surrounding matrix. Notably, this does not preclude full population inversion: because the Rabi frequency scales as $\sqrt{I}$ while the ESA rate scales as $I$, pulsed resonant excitation shorter than the excited-state lifetime can drive the molecule to inversion while keeping ESA effects negligible. More broadly, our results show that the optical response of DBT molecules under strong driving cannot always be reduced to an effective two-level picture, and that accounting for excitation-induced depletion channels will be essential for future experiments operating deep in the nonlinear regime, from single-photon sources to molecular optomechanics.

{\it Acknowledgments.}---
The project has been co-funded by the European Union (ERC, QUINTESSEnCE, 101088394). Views and opinions expressed are however those of the author(s) only and do not necessarily reflect those of the European Union or the European Research Council. Neither the European Union nor the granting authority can be held responsible for them. The authors wish to thank Z. S. Sadeq and J. E. Sipe for the useful discussions on the theoretical absorption properties of DBT, and D. Martin-Cano and S. Go\"{e}tzinger for insightful discussions. 

\clearpage
\onecolumngrid

\setcounter{secnumdepth}{2}
\renewcommand{\thesection}{S\arabic{section}}
\renewcommand{\thesubsection}{\thesection.\arabic{subsection}}

\makeatletter
\renewcommand{\p@subsection}{}
\makeatother

\setcounter{section}{0}

\startcontents[appendix]
\renewcommand{\thefigure}{S\arabic{figure}}
\setcounter{figure}{0}
\begin{center}
\textbf{\Large Supplemental Material for
\\``Breakdown of the optical saturation regime in  molecular single-photon emitters''}
\bigskip

Hugo Levy-Falk$^1$
Daniele De Bernardis$^{1,2}$
Elena Fanella$^{3}$
Louise Morlaes$^{1,2}$
Costanza Toninelli$^{1,2}$
toninelli@lens.unifi.it
hugoloup.levy-falk@ino.cnr.it

\bigskip

$^{1}${\it National Institute of Optics (CNR-INO), c/o LENS via Nello Carrara 1, Sesto F.no 50019, Italy}

$^{2}${\it European Laboratory for Non-Linear Spectroscopy (LENS), Via Nello Carrara 1, Sesto F.no 50019, Italy}

$^{3}${\it Physics Department, University of Naples, Via Cinthia 21, Fuorigrotta 80126, Italy}

\bigskip

In this Supplemental Material, we provide additional information on experimental methods used. Additional figures and data from the experiment giving a larger pool of examples. Additional experimental results supporting the ruling out of thermal effects. The theory of thermal effects via phononic coupling between the molecule and the matrix. The basic theory of a molecule as a two-level emitter. The theory of the two-level molecule with the additional dissipative dark-state included, giving rise to the excited state absorption.
Additional experimental data and figures regarding intersystem crossing. Additional data and figures supporting a possible correlation with frequency shift due to optical pumping.
\end{center}
\maketitle
\tableofcontents
\newpage




\section*{Notation and symbols}
\label{sec:tables}

\begin{table}[h!]
    \centering
    \begin{tabular}{ccc}
        \hline
         {\bf Symbol} & {\bf Typical value (or range)} & {\bf Parameter name}   \\
         \hline
         \hline
        $I$ & $[10^{-1};10^{7}]\,{\rm W/cm}^2$ & Laser intensity \\
         \hline
        $I_{\rm sat}^{\rm res}$ & $[1;10]\, {\rm W/cm}^2$ & Saturation intensity (resonant excitation)\\
         \hline
        $R_{\infty}^{\rm res}$ & $[10;10^2]$kcount/s & Asymptotic fluorescence rate (resonant excitation)\\
        \hline
        $I_{\rm sat}^{\rm off}$ & $[10^2;10^3]\, {\rm W/cm}^2$ & Saturation intensity (off-resonant excitation)\\
         \hline
        $R_{\infty}^{\rm off}$ & $[10^2;10^3]$kcount/s & Asymptotic fluorescence rate (off-resonant excitation)\\
        \hline
        $S$ & [0;100] & Saturation parameter \\
        \hline
        $S^{\rm res}$ & [0;100] & Saturation parameter (specific for resonant excitation)\\
        \hline
        $S^{\rm off}$ & [0;1] & Saturation parameter (specific off-resonant excitation)\\
        \hline
        $\xi^{\rm res}$ & [$10^{-3};10^{-2}$] & Anomaly parameter (resonant excitation)\\
        \hline
        $\xi^{\rm off}$ & [$10^{-1};1$] & Anomaly parameter (off-resonant excitation)\\
        \hline
        $I_{\rm ESA}$ & $[10^2;10^3]$ W/cm$^2$ & ESA activation intensity\\
        \hline
        $\gamma_{\rm ESA}$ & $2\pi[0;10]\,$GHz & ESA effective dissipation rate\\
        \hline
        \hline
        $\gamma_1$ & $2\pi\times$[40;100] MHz & Spontaneous emission rate (inverse lifetime)\\
        \hline
        $\gamma_{\phi}$ & $2\pi\times$[0;10] GHz & Dephasing rate \\
        \hline
        $\gamma_{2}$ & $2\pi\times$[$10^{-2}$;10] GHz & Decoherence rate (inverse coherence time) \\
        \hline
         $\Gamma_{\rm zpl}$ & $2\pi\times$[$10^{-2}$;10] GHz & ZPL's Full Width Half Maximum\\
        \hline
        $\gamma_{\rm v}$ & $2\pi\times$[10;20] GHz & Vibrational inverse lifetime \\
        \hline
        $\gamma_{\rm rad}$ & $2\pi\times$[40;80] MHz & Radiative decay rate\\
        \hline
        $\Gamma_+$ & $2\pi\times$[0;10] GHz & Incoherent pump rate\\
        \hline
        $\Omega_0$ & $2\pi\times$[0;10] GHz & Bare Rabi frequency \\
        \hline
        \hline
        $T$ & $[3;30]\,$K & Temperature \\
        \hline
        $F$ & [0;0.1] & Huang-Rhys factor \\
        \hline
        $\Omega_T$ & $2\pi\times$[0;10] GHz & Thermal Rabi frequency\\
        \hline
        $\omega_{\phi}$ & $2\pi\times$[0;10] GHz & Thermal frequency shift \\
        \hline
        $T_{\omega}$ & $[1;30]\,$K & Thermal frequency-shift activation temperature\\
        \hline
        $T_{\phi}$ & $[1;30]\,$K & Thermal dephasing activation temperature\\
        \hline
        $T_{F}$ & $[1;30]\,$K & Thermal Huang-Rhys factor activation temperature\\
        \hline
        $a,b$ & $[0;10]$ & Thermal exponents \\
        
        \hline
    \end{tabular}
    \caption{Main physical quantities used throughout the manuscript, typical values, and names.}
    \label{tab:phys_quantities}
\end{table}

All the quantities marked with $\tilde{\ldots}$ (\emph{tilde}) refer to quantities in Table \ref{tab:phys_quantities} modified by the anomalous saturation (both experimental or theory defined quantities).

\begin{table}[h!]
    \centering
    \begin{tabular}{cc}
        \hline
         ZPL & Zero-Phonon Line \\
         \hline
         DBT & Dibenzoterrylene \\
         \hline
         TLS & Two-Level System \\
         \hline
         CW & Continuous Wave \\
         \hline
         NC & Nanocrystal \\
         \hline
         DBT:Ac & DBT embedded in Ac crystal \\
         \hline
         FWHM & Full Width at Half Maximum \\
         \hline
         ESA & Excited States Absorption \\
         \hline
         DW & Debye-Waller (factor)\\
         \hline
         ISC & Intersystem Crossing \\
         \hline
         SM & Supplemental Material\\

        \hline
    \end{tabular}
    \caption{Acronyms}
    \label{tab:acronyms}
\end{table}

\pagebreak

\section{Methods}
\label{sec:methods}
\subsection{Sample preparation}
The DBT-doped Ac nanocrystals were fabricated using a reprecipitation method. A \SI{1}{m\Molar} solution of DBT in toluene was mixed with a \SI{5}{m\Molar} solution of Ac in acetone, obtaining a final solution of DBT:Ac in acetone with a concentration of \SI{100}{n\Molar}. The mixture was then sonicated for \SI{30}{min} to allow the evaporation of toluene and the simultaneous crystallization of Ac around the DBT molecules. The resulting suspension of DBT-doped Ac nanocrystals in acetone was deposited on a gold-coated glass substrate. The solvent was subsequently evaporated under vacuum, leaving the nanocrystals alone on the substrate.

The microcrystals are fabricated via a sublimation process. In a vial equipped with a shutter and a filtering system we insert \SI{80}{\micro L} of a \SI{1}{m\Molar} solution made of DBT diluted in Toluene and \SI{144}{mg} of Ac in powder~\cite{Lange2026}. After three cycles of cleaning the atmosphere by pumping nitrogen and vacuum alternatingly, we set the pressure to \SI{800}{mbar} and start heating the environment with a heat gun, placed below the vial, at \SI{380}{\degree C}. When the mixture melts, we open the shutter and continue the heating until the microcrystals are transported through the filtering system and collected on the gold-glass substrate positioned at the top of the vial.

\subsection{Optical setup}
\label{sec:optical setup}

\begin{figure}
    \centering
    \includegraphics[width=0.5\columnwidth]{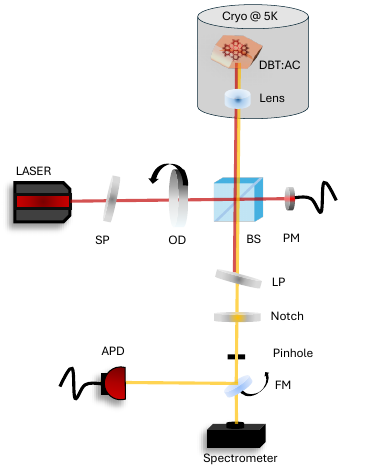}
    \caption{Schematic of the set-up used in the experiments. LASER: exciting beam. SP: short-pass filter. OD: rotating optical density. BS: beam splitter 30:70 (Reflection : Transmission). PM: power meter. CRYO: cryostat. LP: long-pass filter. Notch: sharp band-pass filter centered around the ZPL. FM: flipable mirror. APD: avalance photo-diode.}
    \label{fig:setup}
\end{figure}

Fig.~\ref{fig:setup} shows the experimental set-up used for both the resonant and the off-resonant measurements. The excitation is performed in the resonant case by either a DFB laser \textsc{Toptica LD-0785-0080-DFB-1} or by a \textsc{Syrah Matisse C} continuous-wave laser, while in the off-resonant case we use the Matisse or the \textsc{Toptica DXL 110}. A short-pass filter, either \textsc{semrock tsp01-790} or \textsc{semrock FF01-769/41-25}, is positioned to cut the long-wavelength emission of the laser. The power is adjusted by means of neutral density filter wheel. After a 30:70 (R:T) beam splitter \textsc{BS020}, a fraction of the excitation beam is used to monitor the incident laser intensity with a power meter. The reflected laser light is focused on the sample mounted inside a cryostat at \SI{4}{K} (either \textsc{Montana Cryostat} for the resonant case or \textsc{Entropy Cryostat} for the off-resonant experiment), through a microscope objective \textsc{NA = 0.67, SigmaKoki PAL-50-NIR-HR-LC07} or a 0.7-NA focusing lens, respectively. The sample is excited and the fluorescence emitted by the molecule is collected through the same system and spatially filtered with a pinhole in a confocal configuration. In the resonant excitation scheme, we use a long-pass filter \textsc{Semrock Razoredge-785RS-25} while in the off-resonant case we combine a long-pass filter \textsc{Thorlabs FELH0750} with an additional notch filter (\textsc{optigrate bnf-785-od4-12-5m}) centered at \SI{785}{nm} with a \SI{0.4}{nm} bandwidth isolates a narrow spectral window around the ZPL. The fluorescence signal is directed by a flip mirror either to an avalanche photodiode for photon counting  \textsc{
Excelitas SPCM 800 14 FC} or to a spectrometer  \textsc{andor sr303i-A} for spectral measurements.

The off-resonant saturation curves shown in Fig.\ref{fig:off-resonant saturation}, are measured by recording simultaneously the laser power and the integrated fluorescence of the molecule, selected with the notch filter, on the APD, while the laser power is being continuously varied. The saturation curve is then reconstructed by matching the timestamps.

Resonant saturation experiments, such as shown in Fig.~\ref{fig:2}, are performed by integrating the Stokes-shifted fluorescence of the molecule while the laser scans the frequencies around the ZPL.

\subsection{Data acquisition and analysis}

Our custom optical microscopy setup leverages the Qudi framework~\cite{binderQudiModularPython2017} for automation. 

The data analysis of the experimental data was performed with the Julia programming language~\cite{bezansonJuliaFreshApproach2017}, leveraging its rich ecosystem~\cite{datserisDrWatsonPerfectSidekick2020, Bouchet_Valat2023}. The plots presented here were produced using the Makie framework~\cite{DanischKrumbiegel2021}.

\subsection{Simulations}
The simulations presented in this work were performed using the Qutip framework~\cite{Lambert2026}.

\section{Supplementary data in the anomalous saturation regime}
\subsection{Maximum achievable saturation parameter and excited state population}
\label{sec:S_param_appendix}
\begin{figure}
    \centering
    \begin{subcaptiongroup}
        \subcaptionlistentry{S parameter}
        \label{fig:S parameter}
        \subcaptionlistentry{population extraction}
        \label{fig:population extraction}
    \end{subcaptiongroup}
    \includegraphics[width=0.6\linewidth]{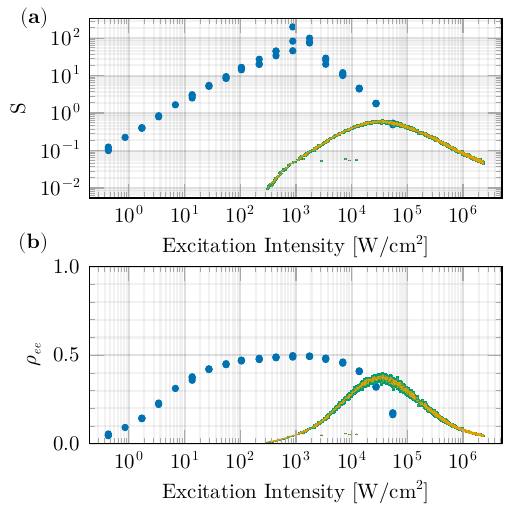}
    \caption{\subref{fig:S parameter}~Saturation parameter $S$ as a function of excitation laser intensity for resonant excitation (blue dots) and off-resonant excitation (yellow dots). \subref{fig:population extraction}~Corresponding population of the excited state ($\rho_{ee}$). In both sub-figures, the green bars indicate the uncertainty on the reported values in the off-resonant case.}
    \label{fig:4}
\end{figure}

The fluorescence measurements presented in the main text, employing different pumping schemes, can be compared by estimating the saturation parameter, obtained by inverting Eq.~\eqref{eq:standard saturation}, which yields $S = R(I)/(R_{\infty} - R(I))$. For $0-0$ excitation, the fluorescence rate $R^{\rm res}(I)$ is directly measurable, while the asymptotic rate is approximated as $R_{\infty}^{\rm res} \approx \max_{I} R^{\rm res}(I)$, since a plateau is clearly visible (Fig.~\ref{fig:resonant saturation  amplitude}); this implicitly assumes that the excited-state population reaches its TLS maximum $\rho_{ee}\approx0.5$ under resonant CW excitation (Sec.~\ref{sec:OBE}). The extracted $S^{\rm res}$ is shown as blue dots in Fig.~\ref{fig:S parameter}, reaching a maximum value $\approx100$ -- a quantity that would instead grow indefinitely with intensity for a simple TLS.

For the $0-1$ off-resonant scheme, no clear plateau is visible in Fig.~\ref{fig:off-resonant saturation}, making $R_{\infty}^{\rm off}$ and hence $S^{\rm off}$ ill-defined from a direct fit. Using the ESA-based fitting procedure described in Sec.~\ref{sec:uni_offres_AS} instead yields the yellow dots in Fig.~\ref{fig:S parameter}, with $\max_{I} S^{\rm off}\in[0.1;1]$ -- much smaller than the resonant case. Despite the higher absolute ZPL count rate under off-resonant pumping, the molecule is therefore not brought above the saturation threshold ($S=1$), indicating a small excited-state population $\rho_{ee}\lesssim0.5$ (Fig.~\ref{fig:population extraction}, yellow dots, compared to the resonant case in blue).

\subsection{Universality of the anomalous saturation behavior}
\label{sec:uni_offres_AS}
\begin{figure}[h!]
    \centering
    \begin{subcaptiongroup}
        \subcaptionlistentry{off-resonant saturation amplitude}
        \label{fig:sup:off-resonant saturation amplitude}
        \subcaptionlistentry{emission spectrum}
        \label{fig:sup:emission spectrum}
    \end{subcaptiongroup}
    \includegraphics[width=\textwidth]{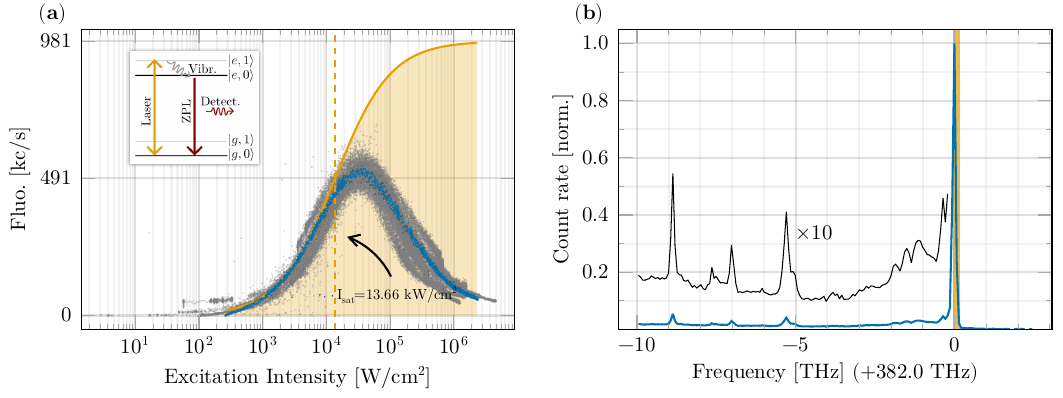}
    \caption{\subref{fig:sup:off-resonant saturation amplitude}~Figure reproduced from Fig.~\ref{fig:off-resonant saturation}. The gray dots report the saturation curves for 18 other molecules, scaled in intensity to account for the variabilities in efficiency of excitation and in count rate to account for the variabilities in efficiency of detection. \subref{fig:sup:emission spectrum}~Typical emission spectrum of DBT molecules when excited off-resonantly (inset in \subref{fig:sup:off-resonant saturation amplitude}). The black line shows transmission of the selective filter, the orange vertical span indicates the spectral range that was recorded in \subref{fig:sup:off-resonant saturation amplitude}. Since it is used in reflection (see 
Sec.~\ref{sec:optical setup}), the dip gives the spectral region integrated to get the curve in \subref{fig:sup:off-resonant saturation amplitude}.}
    \label{fig:off-resonant saturation bis}
\end{figure}

\begin{figure}
    \centering
    \begin{subcaptiongroup}
        \subcaptionlistentry{average saturation intensity off-resonant}
        \label{fig:sup:average saturation intensity off-resonant}
        \subcaptionlistentry{average R infty off-resonant}
        \label{fig:sup:average R infty off-resonant}
        \subcaptionlistentry{average xi off-resonant}
        \label{fig:sup:average xi off-resonant}
        \subcaptionlistentry{example anomalous saturation fit}
        \label{fig:sup:example anomalous saturation fit}
    \end{subcaptiongroup}
    \includegraphics[width=\textwidth]{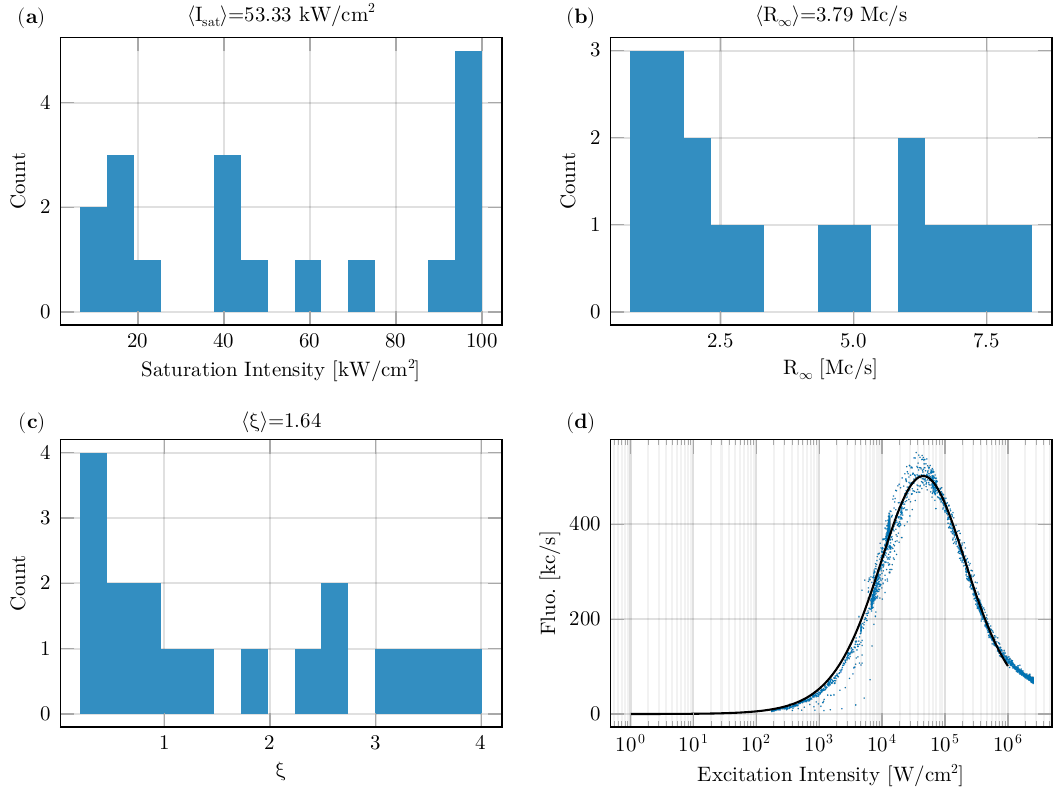}
    \caption{Fitted parameters of Equation~\ref{eq:anomalous_saturation} over the 18 off-resonant saturation curves shown in~\ref{fig:sup:off-resonant saturation amplitude}: \subref{fig:sup:average saturation intensity off-resonant}~Saturation intensity, \subref{fig:sup:average R infty off-resonant}~$R_\infty$, and \subref{fig:sup:average xi off-resonant}~$\xi$. \subref{fig:sup:example anomalous saturation fit}~Representative example of the quality of the fit (solid black line) for one of the saturation curves (blue dots).}
    \label{fig:sup:anomalous saturation fir off-resonant}
\end{figure}

The anomalous saturation behavior is systematically observed for DBT in anthracene nanocrystals. Supl.~Fig.~\ref{fig:sup:off-resonant saturation amplitude} shows the data of Fig.~\ref{fig:off-resonant saturation} superimposed with the saturation curves from 18 other molecules (in gray dots). It is clear from the picture that the behavior is reproducible across different molecules. In complement, Supl.~Fig.~\ref{fig:sup:anomalous saturation fir off-resonant} shows that fitting Equation~\ref{eq:anomalous_saturation} on this dataset yields homogeneous results, with respective average values of $I_\text{sat}$, $R_\infty$, and $\xi^\text{off}$ of \SI{53}{kW/cm^2}, \SI{3.79}{Mc/s}, and \qty{1.64}.

\begin{figure}
    \centering
    \begin{subcaptiongroup}
        \subcaptionlistentry{prior off-resonant}
        \label{fig:sup:prior off-resonant}
        \subcaptionlistentry{posterior off-resonant}
        \label{fig:sup:posterior off-resonant}
    \end{subcaptiongroup}
    \includegraphics[width=\textwidth]{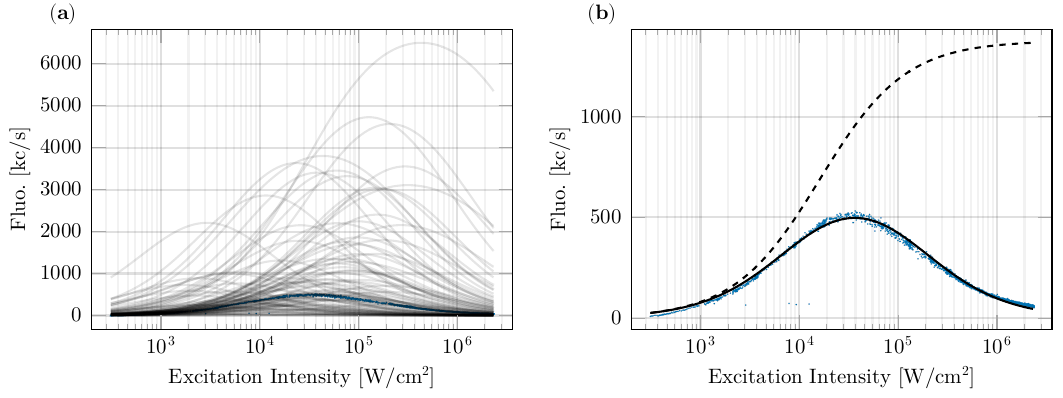}
    \caption{\subref{fig:sup:prior off-resonant}~Sample of 1000 anomalous saturation curves from the prior distribution (solid light gray curves) for the data of Fig.~\ref{fig:off-resonant saturation} of the main text (blue dots). \subref{fig:sup:posterior off-resonant}~Sample of 4000 anomalous saturation curves from the posterior distribution (solid gray lines) for the data of Fig.~\ref{fig:off-resonant saturation} of the main text (blue dots). The dashed black line corresponds to the equivalent normal saturation curve ($\xi$ set to 0).}
    \label{fig:sup:anomalous saturation off-resonant prior and posterior}
\end{figure}

The classical saturation model shown in Fig.~\ref{fig:off-resonant saturation} was obtained by fitting the anomalous saturation model described in Equation~\ref{eq:anomalous_saturation} using the Turing.jl library~\cite{10.1145/3711897, pmlr-v84-ge18b} and then forcing the parameter $\xi$ to be zero. In ~Fig.~\ref{fig:sup:anomalous saturation off-resonant prior and posterior}, we show the details of the Bayesian fitting procedure of the model defined as:
\begin{align}
    I_\text{sat} &\sim \text{Uniform}(10^3, 100\times10^3)\\
    R_\infty &\sim \text{Uniform}(100\times10^3, 10\times10^6)\\
    \xi &\sim \text{LogNormal}(0, 1)\\
    S_i &= \text{intensity}_i / I_\text{sat}\\
    \sigma &\sim \text{LogNormal}(0, 1)\\
    \mu_i &= R_\infty * S / ((1+\xi*S)^2 + S)\\
    \text{count}_i &\sim \text{Normal}(\mu_i, \sigma)
\end{align}
where \texttt{count} and \texttt{intensity} are the experimental count rates and intensities. In particular, Supl.~Fig.~\ref{fig:sup:prior off-resonant} shows a sample of 1000 curves from the the prior distribution, while Supl.~Fig.~\ref{fig:sup:posterior off-resonant} shows a sample of 4000 curves from the posterior distribution (solid lines).

\begin{figure}
    \centering
    \includegraphics{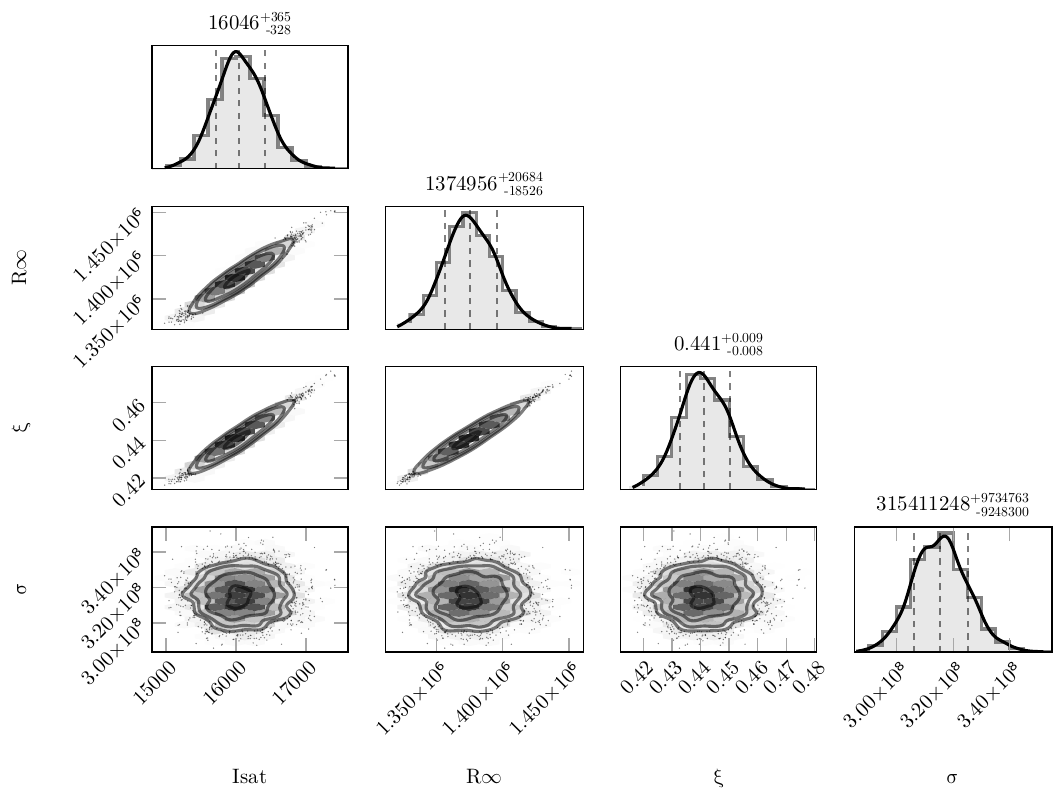}
    \caption{Parameter distribution of the posterior distribution when fitting the dataset of Fig.~\ref{fig:off-resonant saturation} of the main text with the anomalous saturation model. $R_\infty$ is in $\text{c}/\text{s}$, $\sigma$ in $(\text{c}/\text{s})^2$, and $I_\text{sat}$ in $\text{W}/\text{cm}^2$.}
    \label{fig:sup:off-resonant pairplot}
\end{figure}

In~Fig.~\ref{fig:sup:off-resonant pairplot}, we show the posterior distribution of parameters. While it is to be noted that there is some cross-dependency on the fitted parameters, we highlight that the relative spread of the samples is reasonably small. Theses sampled parameters were then used to generate the off-resonant curve in Fig.~\ref{fig:4} above.

\subsection{Stokes-shifted fluorescence in the anomalous regime}
\label{sec:stokes_shifted}

\begin{figure}[h!]
    \centering
    \includegraphics{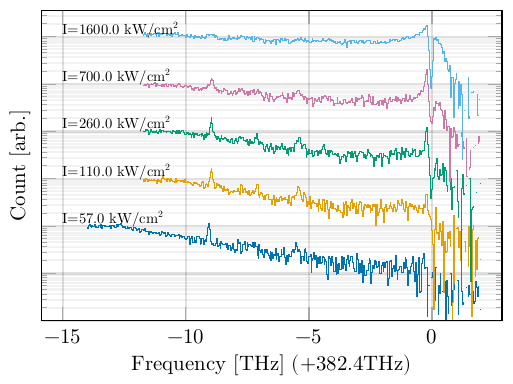}
    \caption{Emission spectra in the anomalous saturation regime. The Zero-Phonon Line has been filtered out using a narrow Notch filter to see only the vibrational peaks. This yields the dip visible at zero frequency. The signal from each successive spectrum has been multiplied by ten to improve legibility.}
    \label{fig:sup:emission spectra amomalous saturation}
\end{figure}

To rule-out the possibility that the anomalous-saturation could be explained by some additional light being emitted on the Stokes-shifted peaks, we recorded emission spectra while exciting off-resonantly the molecule (as in the inset of~\ref{fig:sup:off-resonant saturation amplitude}) for increasing excitation intensities. This experiment is reported in Supl.~Fig.~\ref{fig:sup:emission spectra amomalous saturation}. It is clear from the plot that there is no significant increase of the fluorescence of the molecule.

\begin{figure}
    \centering
    \begin{subcaptiongroup}
        \subcaptionlistentry{off-resonant saturation everything}
        \label{fig:sup:off-resonant saturation everything}
        \subcaptionlistentry{off-resonant saturation stokes-shifted}
        \label{fig:sup:off-resonant saturation stokes-shifted}
    \end{subcaptiongroup}
    \includegraphics[width=\linewidth]{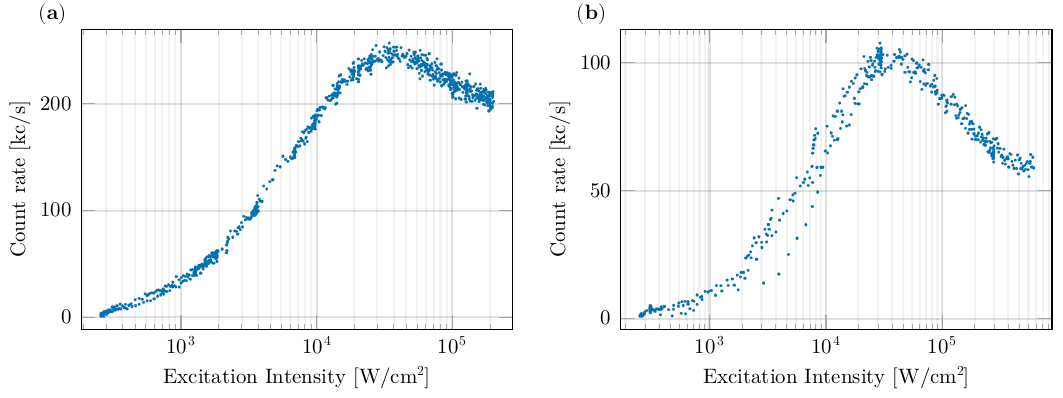}
    \caption{\subref{fig:sup:off-resonant saturation everything}~Saturation curve recorded in the off-resonant configuration when the whole emission spectrum is integrated.  \subref{fig:sup:off-resonant saturation stokes-shifted}~Saturation curve recorded in the off-resonant configuration when ony the Stokes-shifted emission is integrated (with the ZPL excluded). }
    \label{fig:sup:off-resonant saturation other areas}
\end{figure}

In complement, one can record a saturation curve such as the one presented in Fig.~\ref{fig:off-resonant saturation}, but this time integrating the whole emission spectrum (Supl.~Fig.~\ref{fig:sup:off-resonant saturation everything}) and only the Stokes-shifted fluorescence (Supl.~Fig.~\ref{fig:sup:off-resonant saturation stokes-shifted}). It is clear from both these pictures that the anomalous saturation can be observed for the corresponding experimental configurations. 

\FloatBarrier
\pagebreak
\subsection{Reversibility of the anomalous saturation}
\label{sec:reversibility}

\begin{figure}[h!]
    \centering
    \includegraphics{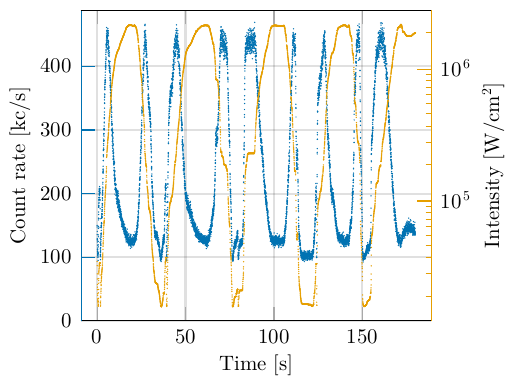}
    \caption{Time-dependent record of the emission rate from a crystal excited off-resonantly (blue dots) while the excitation laser intensity is being modulated (yellow dots). The anomalous saturation regime is clearly seen at high intensity, and it is also clear that the molecule recovers from its emission being quenched.}
    \label{fig:sup:anomalous saturation reversibility}
\end{figure}

The first explanation that comes to mind for any behavior involving a reduction of fluorescence at high laser intensity of a molecular emitter is photobleaching, \textit{i.e.} denaturation of the molecule. However, this can be easily disproved by noting the reversibility of the process. Fig.~\ref{fig:sup:anomalous saturation reversibility} shows the level of fluorescence emitted at the wavelength of the ZPL of a molecule excited off-resonantly (blue dots) while the laser intensity is being scanned (yellow dots). The anomalous saturation regime, corresponding to high laser intensity but low fluorescence level, is reached five times during this scan. It is clear that the molecule recovers the normal saturation regime in between these anomalous saturation phases, with the level of fluorescence being positively correlated with the laser intensity.

\FloatBarrier
\pagebreak
\subsection{Resonant anomalous saturation regime}

\begin{figure}[h!]
    \centering
    \includegraphics{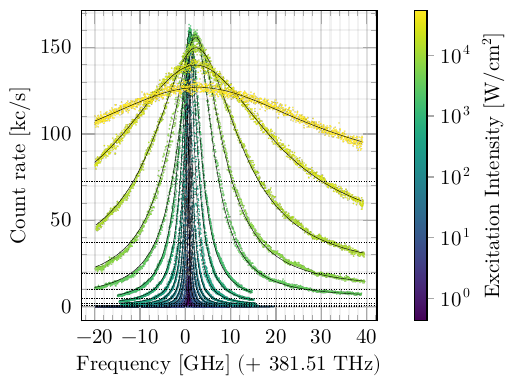}
    \caption{Raw data for the saturation curve shown in Fig.~\ref{fig:2} of the main text. It should be noted that these measurements include a power-dependent offset due to the laser leaking in the detection part of the setup.}
    \label{fig:sup:resonant scans anomalous saturation}
\end{figure}

As shown in the main text, the anomalous saturation regime is also observed in the case of resonant excitation. In this configuration, a laser scans the ZPL and we record the Stokes-shifted fluorescence (see the inset in Fig.~\ref{fig:resonant saturation scan}). As expected for a two-levels system, one gets a lorentzian line shape~\cite{loudonQuantumTheoryLight2000}. Supl.~Fig.~\ref{fig:sup:resonant scans anomalous saturation} shows how we fit such scans (solid black lines). We note that the anomalous saturation is already visible here, as the curve corresponding to the highest excitation intensity (yellow) lies below the ones corresponding to mid-range excitation intensities (green). However, the effect is even more apparent when one removes the signal baseline (dashed horizontal lines). This baseline accounts for a significant part of the signal at high laser intensity because the laser leaks in the detection setup. The main text uses the corrected dataset.

\FloatBarrier
\pagebreak
\section{Ruling out laser-induced heating}
\label{sec:heating_ruled_out}
\subsection{Experimental Results}
\begin{figure}
    \centering
    \begin{subcaptiongroup}
        \subcaptionlistentry{temperature scan broadening}
        \label{fig:temperature scan broadening}
        \subcaptionlistentry{laser heating}
        \label{fig:laser heating}
        \subcaptionlistentry{comparison DW factor temperature and intensity scan}
        \label{fig:comparison DW factor temperature and intensity scan}
    \end{subcaptiongroup}
    \includegraphics[width=0.6\linewidth]{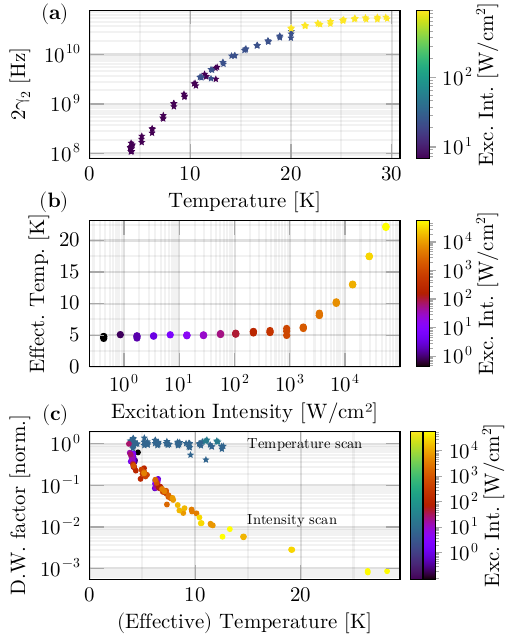}
    \caption{\subref{fig:temperature scan broadening}~Power-broadening-correct linewidth of the ZPL transition of a single DBT molecule as a function of temperature. The interpolation of that dataset gives a law linking linewidth to temperature. \subref{fig:laser heating}~Effective temperature of a molecule undergoing the anomalous saturation effect when interpolating the excess of linewidth with the temperature law of~\ref{fig:temperature scan broadening}. \subref{fig:comparison DW factor temperature and intensity scan}~Comparison of the Debye-Waller factor as a function of the cryostat temperature (blue-ish stars) and as a function of the effective temperature of a molecule experiencing an anomalous saturation. }
    \label{fig:5}
\end{figure}

As discussed in the main text, we consider the anomalous saturation as a result of potential heating, due to the intense laser light impinging on the sample.
While the Ac matrix is transparent at the laser wavelength around \SI{785}{nm}, the gold substrate could e.g. warm up, absorbing beyond the cooling capabilities of our cryostat. In this case, the molecule would sense an effective temperature, $T_{\rm eff}(I)$, monotonically increasing with the laser power. Following an approach similar to the one presented in Ref.~\cite{clear_phonon-induced_2020}, the temperature effect on the molecule optical properties is estimated (see Sec.~\ref{sec:thermal_theory}).
In particular, the optomechanical coupling between the molecule's electronic energy levels and the matrix phonons, can be summarized in the following effects:
\begin{itemize}
    \item a thermal increase of the decoherence rate $\gamma_2(T)=(\gamma_1 + \gamma_{\phi}(T))/2$, due to phonon-induced dephasing rate $\gamma_{\phi}(T)$~\cite{clear_phonon-induced_2020};
    \item a reduction of the optical cross-section through the so-called \emph{Huang-Rhys factor}, $F(T)>0$, effectively reducing the molecule's Rabi frequency as $\Omega_{T} = \Omega_0e^{-F(T)}$, with $e^{-F(T)}$ being equivalent to the Debye-Waller factor (DW)\cite{clear_phonon-induced_2020};
    \item a thermal frequency shift $\omega_{\phi}(T)$ linked to the appearence of the phonon-induced dephasing~\cite{Mirzaei2026}.
    \item a modified emission spectrum where the relative weight between the ZPLs and the phonon wing is accounted for by the DW factor.
\end{itemize}

Altogether, these phenomena determine a temperature-dependent emission rate and therefore saturation parameter, according to Eq.~\eqref{eq:S_param_thermal}, where the temperature dependence of the Huang-Rhys factor $F(T)$ and of the linewidth are discussed in Sec.~\ref{sec:thermal_theory} and we used $2\gamma_2(T)=\gamma_1 + \gamma_{\phi}(T)$ to highlight the explicit dependence on the pure dephasing ratio $\gamma_{\phi}(T)/\gamma_1$.

In the following we exploit the fact that the thermal contribution to the decoherence rate $\gamma_{\phi}(T)$ yields a direct probe of the environment temperature, as already demonstrated e.g. in Ref.~\cite{estesoQuantumThermometrySingle2023}. In particular, scanning the cryostat temperature, the linewidth broadening can be calibrated, yielding effectively a temperature reading (See Fig.~\ref{fig:temperature scan broadening}).
The linewidth-to-temperature map can  then be used to estimate an effective temperature in the anomalous saturation scan of Fig.~\ref{fig:2}, as a function of the laser intensity $T_{\rm eff}(I)$. The results of this analysis are plotted in Fig.~\ref{fig:laser heating}.

Now, investigating the hypothesis of laser-induced heating, the fluorescence reduction as a function of temperature is compared for the case of direct increase via cryostat heaters and for the one of growing laser power. Particularly striking is the difference in the DW factor, as extracted using Eq.~\eqref{eq:S_param_thermal}. The two cases are plotted together in Fig.~\ref{fig:comparison DW factor temperature and intensity scan}, showing orders of magnitude discrepancy in the scaling with temperature. 
It is worth noticing that the measured DW factor behavior in the temperature scan is consistent with what is reported, e.g., in Ref.~\cite{clear_phonon-induced_2020}.
The relatively smooth temperature dependence of the DW factor has an even stronger implication, looking at the off-resonant experiment. 
In this case, thermal dephasing does not affect the off-resonant saturation parameter, which becomes solely dependent on temperature through the DW factor (see Sec.~\ref{sec:OBE_off_res}). Hence, the pronounced depletion of the emission rate under off-resonant pumping can hardly be described by a temperature effect.

In conclusion, the hypothesis of a simple laser-induced heating can be generally ruled out.

\section{Theory: the influence of thermal phonons}
\label{sec:thermal_theory}
We consider here a single molecule that is mechanically coupled to a phononic bath.
For our aims, to a first approximation, the molecule can be described as a two-level system and a single harmonic vibrational mode.
The bath of phonons is described as a collection of $n=1,2\ldots$ harmonic oscillators, characterised by a certain dispersion $\omega_n$.
Using the approach of Refs.~\cite{clear_phonon-induced_2020,De_Bernardis2025}, the total Hamiltonian in the multi-mode polaron frame is
\begin{equation}
    H = H_{\rm mol} + H_{\rm phon} + H_{\rm mol-phon} + H_{\rm drive}(t),
\end{equation}
where
\begin{equation}\label{eq:Hmol_0}
    H_{\rm mol} = \hbar\omega_0\hat{\sigma}^{\dag}\hat{\sigma} + \hbar\omega_{\rm v}\hat{b}^{\dag}\hat{b},
\end{equation}
\begin{equation}\label{eq:Hphonon_discrete}
    H_{\rm phon} = \sum_n \hbar \omega_n \hat{b}^{\dag}_n\hat{b}_n,
\end{equation}
\begin{equation}\label{eq:Hmol-phonon_int_discrete}
    H_{\rm mol-phon} = \sum_n \hbar \chi_n\, \hat{b}^{\dag}\hat{b}_n \hat{\sigma}^{\dag}\hat{\sigma},
\end{equation}
\begin{equation}\label{eq:Hdrive_discrete}
    H_{\rm drive}(t) = \hbar \Omega_0 \left[ \hat{\sigma}e^{-\eta_{\rm v}(\hat{b}-\hat{b}^{\dag}) - \sum_n\eta_n(\hat{b}_n-\hat{b}^{\dag}_n) + i\omega_Lt} + {\rm H.c.} \right].
\end{equation}
Here we have neglected the phonon-squeezing term~\cite{De_Bernardis2025, zhang_nonlinear_2024}, and we have introduced the following parameters:
\begin{itemize}
    \item The HOMO-LUMO electronic transition frequency $\omega_0$, the vibrational frequency (of a single specific mode) $\omega_{\rm v}$, and the matrix's phononic dispersion $\omega_n$.
    \item The phononic parabolic asymmetry $\chi_n$. This parameter is due to having different phononic degrees of freedom acting on the HOMO-PES and LUMO-PES~\cite{De_Bernardis2025, zhang_nonlinear_2024}. It regulates the amount of pure dephasing, as will be clear in the following. It can be equivalently interpreted as a second-order contribution from the phononic strain interaction~\cite{clear_phonon-induced_2020}.
    \item The external laser bare Rabi frequency $\Omega_0$, together with the external laser frequency $\omega_L$.
    \item The vibrational Franck-Condon factor $\eta_{\rm v}$~\cite{De_Bernardis2025}, and the equivalent normalized mechanical coupling for the phonons $\eta_n$.
\end{itemize}
We remind that the bare Rabi frequency depends only on the laser intensity, fundamental constants, and the molecule's transition dipole moment, reading
\begin{equation}\label{eq:Rabi_bare}
    \Omega_0 = d_{\rm DBT}\sqrt{\frac{8\pi\alpha_{\rm fs}}{e^2\hbar} I}.
\end{equation}
Here $d_{\rm DBT} = \langle {\rm HOMO} | \hat{d} | {\rm LUMO}\rangle$ is the DBT dipole matrix element (measured in \emph{Debyes} [D]) of the dipole operator $\hat{d}$ computed between the $|{\rm HOMO}\rangle$ and $|{\rm LUMO}\rangle$ molecular states. 
$\alpha_{\rm fs}\approx 1/137$ is the fine-structure constant, $e\approx 48.26\,$D nm$^{-1}$ is the electron charge, and $\hbar\approx 1.05\times 10^{-4}\,$W/THz$^2$ is the reduced Planck constant.

Since the matrix's phonons are weakly coupled and very dissipative, they only contribute as a dissipation channel for the molecule.
These degrees of freedom constitute a set of virtual levels that are never effectively involved in the dynamics, and can be safely traced away.
In this way, the phonons will enter only in the master equation description of the molecule dynamics.
We first start by treating the effect of $H_{\rm mol-phon}$, which generates the pure-dephasing described in Ref.~\cite{clear_phonon-induced_2020}, and then we describe the effect of phonons through the laser driving $H_{\rm drive}$, which accounts for the Debye-Waller factor and the molecule cross-scattering reduction as a function of temperature.

Before proceeding, we introduce a convenient new representation of the phonons in terms of continuous degrees of freedom $\hat{B}_{\omega}$, such that $[\hat{B}_{\omega}, \hat{B}_{\omega '}^{\dag}] = \delta (\omega- \omega')$.
In order to translate Eqs.~\eqref{eq:Hphonon_discrete}-\eqref{eq:Hmol-phonon_int_discrete}-\eqref{eq:Hdrive_discrete} to this new representation we do the following:
we introduce a continuous dimensionless density of state, the density of asymmetry, and the density of mechanical coupling factor given by
\begin{equation}
    J(\omega) = \sum_n \Delta \omega \,\delta(\omega-\omega_n),
\end{equation}
\begin{equation}
    \chi(\omega) = \sum_n \Delta \omega \, \chi_n\,\delta(\omega-\omega_n),
\end{equation}
and
\begin{equation}
    \eta(\omega) = \sum_n \Delta \omega \, \eta_n\,\delta(\omega-\omega_n).
\end{equation}

Here $\Delta \omega$ is the average frequency spacing between the $n$-modes.
We define a set of continuous operators, such that $\hat{B}_{\omega_n} = \hat{b}_n/\sqrt{\Delta \omega}$, and we take the limit $\Delta\omega\rightarrow 0$, with the condition of keeping $J(\omega )$ finite.
The original discrete Hamiltonians are then rewritten as
\begin{equation}\label{eq:H_phonon_continuous}
    H_{\rm phon} = \hbar\int_0^{\infty} d\omega \, \omega J(\omega)\hat{B}_{\omega}^{\dag}\hat{B}_{\omega},
\end{equation}
\begin{equation}\label{eq:H_mol-phon-int_continous}
    H_{\rm mol-phon} =  \hbar\int_0^{\infty} d\omega\, \chi(\omega)J(\omega) \hat{B}_{\omega}^{\dag}\hat{B}_{\omega} \hat{\sigma}^{\dag}\hat{\sigma},
\end{equation}
\begin{equation}\label{eq:Hdrive_continous}
    H_{\rm drive}(t) = \hbar \Omega_0 \left[ \hat{\sigma}e^{-\eta_{\rm v}(\hat{b}-\hat{b}^{\dag}) - \int_0^{\infty} d\omega \eta(\omega)\sqrt{J(\omega)}(\hat{B}_{\omega}-\hat{B}^{\dag}_{\omega}) + i\omega_Lt} + {\rm H.c.} \right].
\end{equation}

\subsection{Thermal frequency shift}

We focus now on the effect of Eq.~\eqref{eq:H_mol-phon-int_continous}.
At first order, if the phonons are all steadily in the vacuum, this interaction does not do anything, and instead is first-order active (in perturbation theory) only when $\langle \hat{B}_{\omega}^{\dag}\hat{B}_{\omega}\rangle \neq 0$.
Let's define $N_{\omega} = \omega \langle \hat{B}_{\omega}^{\dag}\hat{B}_{\omega}\rangle$ and reabsorb it by defining the bath fluctuation operator $\delta \hat{N}_{\omega} = \omega \hat{B}_{\omega}^{\dag}\hat{B}_{\omega} - N_{\omega}$.
The molecular-phonon Hamiltonian in Eq.~\eqref{eq:H_mol-phon-int_continous} is then rewritten as
\begin{equation}\label{eq:Ham_mol-phon-int_shift_deph}
    H_{\rm mol-phon} = H_{\rm shift} + H_{\rm deph}.
\end{equation}
The first term is a thermal shift of the ZPL molecular frequency
\begin{equation}
    H_{\rm shift} = \hbar \omega_{\phi} \hat{\sigma}^{\dag}\hat{\sigma},
\end{equation}
where
\begin{equation}\label{eq:frequency_shift_deph_thermal}
    \omega_{\phi} =  \int_0^{\infty} d\omega \, \chi(\omega)\frac{J(\omega)}{\omega} N_{\omega}.
\end{equation}

\subsection{Pure dephasing}
\label{sec:pure_deph}
The second term emerging from the partition of Eq.~\eqref{eq:Ham_mol-phon-int_shift_deph} is the origin of the phonon-induced dephasing
\begin{equation}\label{eq:H_deph}
    H_{\rm deph} =  \hbar\int_0^{\infty} d\omega\, \chi(\omega)\frac{J(\omega)}{\omega} \delta\hat{N}_{\omega}\, \hat{\sigma}^{\dag}\hat{\sigma}.
\end{equation}
The interpretation of the two terms in Eqs.~\eqref{eq:frequency_shift_deph_thermal}-\eqref{eq:H_deph} is transparent: a finite population of phonons displaces asymetrically the HOMO and LUMO energies, resulting in a shift of the ZPL frequency, while the fluctuations over this finite phononic population, due to the thermal detailed balance, kick the two energy levels asymmetrically, causing dephasing.

Taking Eq.~\eqref{eq:H_deph} as the standard form of dephasing interaction with an empty bath, we can use the standard derivation of master equation~\cite{gardiner_zoller, petruccione} to extract the following Lindbladian dissipator
\begin{equation}
    \partial_t \hat{\rho} = \frac{\gamma_{\phi}}{2}\left[2 \hat{\sigma}^{\dag}\hat{\sigma}\hat{\rho}\hat{\sigma}^{\dag}\hat{\sigma} - \lbrace{\hat{\sigma}^{\dag}\hat{\sigma}, \hat{\rho}\rbrace} \right].
\end{equation}
Here, the dephasing rate is given by
\begin{equation}
    \gamma_{\phi} = \int_0^{\infty} d\omega \frac{\chi^2(\omega)}{\kappa(\omega)}\frac{J^2(\omega)}{\omega} N_{\omega}.
\end{equation}
We have introduced the phononic dissipation rate $\kappa(\omega)$, which enters the equation through the phononic correlator $\mathcal{C}_{\omega\,\omega'}(0) = {\rm Re}[\int d\tau \langle{ \delta\hat{N}_{\omega}(\tau)\delta\hat{N}_{\omega'}^{\dag}(0)\rangle}] = N_{\omega}/\kappa(\omega)\delta(\omega-\omega')$.
Here we have assumed that the phonon number fluctuating dynamics is purely dissipative, $\delta\hat{N}_{\omega}(\tau) = \delta\hat{N}_{\omega} e^{-\kappa(\omega) t}$.

\subsection{Suppression of the scattering cross-section}

Here we focus on the impact of phonons on the driving term, given by the Hamiltonian in Eq.~\eqref{eq:Hdrive_continous}.
Due to the strong off-resonant character of the electron-phonon interaction and the smallness of the Franck-Condon factors $\eta(\omega)$, this term can be treated only to first order perturbation theory (opposite to the dephasing, which is in second order).
This amounts to just take the partial average of the driving Hamiltonian over the phonon density matrix $\hat{\rho}_{\rm phon}$, replacing $H_{\rm drive} \mapsto {\rm Tr}\left[\hat{\rho}_{\rm phon} H_{\rm drive}\right]$.
When the phonon state is a thermal state at temperature $T$, we can actually work out the exact expression, arriving at
\begin{equation}\label{eq:H_DB_drive_continous}
    H_{\rm drive}(t) = \hbar \Omega_T \left[ \hat{\sigma}e^{-\eta_{\rm v}(\hat{b}-\hat{b}^{\dag}) + i\omega_Lt} + {\rm H.c.} \right].
\end{equation}
with the re-normalized, temperature-dependent, Rabi frequency
\begin{equation}\label{eq:Rabi_DW}
    \Omega_T = \Omega_0 \exp\left[-\frac{1}{2}\int_0^{\infty}d\omega\,\eta^2(\omega)J(\omega)(1+2N_{\omega}(T))\right].
\end{equation}
The exponential suppression factor is called the Debye-Waller factor~\cite{clear_phonon-induced_2020}, and it is more conveniently expressed through the so-called Huang-Rhys factor, $F(T)$
\begin{equation}
    F(T) = \frac{1}{2}\int_0^{\infty}d\omega\,\eta^2(\omega)J(\omega)(1+2N_{\omega}(T)).
\end{equation}

\subsection{Polynomial densities approximation}

Here we rewrite the three contributions derived above, namely, the thermal frequency shift $\omega_{\phi}$, the thermal dephasing $\gamma_{\phi}$, and the thermal Debye-Waller factor $e^{-F}$, in their simplest expressions, to have a simple formula that can be easily fit to the experimental data.
To do so, we need to introduce some assumptions to simplify the spectral densities involved to something that can be treated analytically.

Since the density of states and the density of asymmetry must be continuous functions and vanish for negative frequencies, $\omega < 0$, they both must go to zero at $\lim_{\omega\rightarrow 0^+} J(\omega), \chi(\omega) = 0$. 
We then assume that 
\begin{align}\label{eq:DOS_phonons}
        J(\omega) \approx \mathcal{J}\omega^s e^{-\omega/\omega_{\rm cut}} && \chi(\omega) \approx \mathcal{X}\omega^p,
\end{align}
and similarly to the dissipation 
\begin{equation}
\kappa(\omega) = \mathcal{K}\omega^q.    
\end{equation}
The dimensionality of each constant is thus
\begin{align}
    [\mathcal{J}] = 1/{\rm Hz}^{s} && [\mathcal{X}] = 1/{\rm Hz}^{p-1} && [\mathcal{K}] = 1/{\rm Hz}^{q-1}. 
\end{align}
The shift is then rewritten as
\begin{equation}
    \omega_{\phi} =  \mathcal{J}\mathcal{X}\int_0^{\infty} d\omega \, \omega^{a} e^{-\omega/\omega_{\rm cut}}N_{\omega},
\end{equation}
where 
\begin{equation}
    a=s+p-1,
\end{equation}
and the dephasing rate is
\begin{equation}
    \gamma_{\phi} = \frac{(\mathcal{J}\mathcal{X})^2}{\mathcal{K}}\int_0^{\infty} d\omega \omega^{2a+1-q} e^{-2\omega/\omega_{\rm cut}} N_{\omega}.
\end{equation}
For a typical phononic density of states, we have $s=3$, while $p$ is not really known. For simplicity, we then assume $p=1$, which makes the relative asymmetry $\chi(\omega)/\omega$ constant for each mode, making $a=3$.
Also, the value for $q$ is not easy to understand, so we can also assume $q=1$, such that the dissipation per mode is constant.
Moreover, we assume that the phonon population is in thermal equilibrium, such that
\begin{equation}
    N_{\omega}(T) = \frac{1}{e^{\omega/(k_B T)}-1}.
\end{equation}
The shift and the dephasing rate can then be conveniently rewritten in terms of the special functions
\begin{equation}\label{eq:shift_dephs_rate_final}
    \begin{split}
        \omega_{\phi}(T) &=  \mathcal{J}\mathcal{X}\,\left(\frac{k_B T}{\hbar}\right)^{a+1}\sum_{n=1}\frac{\Gamma(a+1)}{\left( n + k_BT/(\hbar \omega_{\rm cut}) \right)^{a+1}}
        \\
        \gamma_{\phi}(T) & = \frac{(\mathcal{J}\mathcal{X})^2}{\mathcal{K}}\,\left(\frac{k_B T}{\hbar}\right)^{2a+1}\sum_{n=1}\frac{\Gamma(2a+1)}{\left( n + 2k_BT/(\hbar \omega_{\rm cut}) \right)^{2a+1}}.
    \end{split}
\end{equation}
Here we have fixed the value $q=1$, while leaving $a$ free, and we have used the integrals
\begin{equation}
        \Gamma(a+1) = \int_0^{\infty}dx\, x^{a} e^{-x} 
\end{equation}
\begin{equation}
        \int_0^{\infty} \, x^a  \frac{e^{-x/x_0}}{e^x-1} =   \sum_{n=0}^{\infty}\int_0^{\infty} dx \, x^a e^{-\left(1/x_0 + n + 1\right)x} = \sum_{n=0}^{\infty} \frac{1}{X^{a+1}(n)} \int_0^{\infty}dx\, x^a e^{-x} ,
\end{equation}
where $X(n) = 1/x_0 + n + 1$.
We remind that for phononic baths it is quite likely to have $a>3$, and a cut-off frequency $\omega_{\rm cut}/(2\pi)\in [10^2;10^3]\,$GHz.
For the specific case reported above, the dimensionality of the free fit-parameters is
\begin{align}
    [\mathcal{J}\mathcal{X}] = 1/{\rm Hz}^{a} && [\mathcal{K}] = {\rm dimensionless} && [\omega_{\rm cut}] = {\rm Hz}
\end{align}

Again, the integral can be performed by introducing some simple assumptions. To have the Franck-Condon factor per mode finite at $\omega=0$, we take $\eta(\omega) = \mathcal{E}\,\omega^r$, with $r>0$.
The dimension of this new Franck-Condon parameter is
\begin{equation}
    [\mathcal{E}] = 1/{\rm Hz}^{r+1/2}.
\end{equation}
Using the density of states defined in Eq.~\eqref{eq:DOS_phonons}, we can perform the two integrals, defining the two functions
\begin{equation}
F_0  = \frac{1}{2}\int_0^{\infty}d\omega\,\eta^2(\omega)J(\omega) = \frac{\mathcal{J}\mathcal{E}^2}{2}\int_0^{\infty}d\omega\, \omega^{b-1}e^{-\omega/\omega_{\rm cut}}  = \omega_{\rm cut}^{b}\frac{\mathcal{J}\mathcal{E}^2}{2} \Gamma(b) 
\end{equation}
\begin{equation}
\Delta F(T) = \int_0^{\infty}d\omega\,\eta^2(\omega)J(\omega)N_{\omega}(T)  =   \mathcal{J}\mathcal{E}^2\int_0^{\infty}d\omega\,\omega^{b-1}\frac{e^{-\omega/\omega_{\rm cut}}}{e^{\omega/(k_B T)}-1} =  \mathcal{J}\mathcal{E}^2\left(\frac{k_B T}{\hbar}\right)^{b}\Gamma(b)\sum_{n=1}^{\infty}\frac{1}{\left(n + k_BT/(\hbar \omega_{\rm cut})\right)^{b}}.
\end{equation}
Here
\begin{equation}
    b=2r+s+1.
\end{equation}

\begin{equation}\label{eq:cross-section_suppression_F}
\begin{split}
    F(T) &= \mathcal{J}\mathcal{E}^2\Gamma(b)\left[\frac{\omega_{\rm cut}^b}{2} + \left(\frac{k_B T}{\hbar}\right)^{b}\sum_{n=1}^{\infty}\frac{1}{\left(n + k_BT/(\hbar \omega_{\rm cut})\right)^{b}}\right]
    \\
    &  = F_0 + \Delta F(T).
\end{split}
\end{equation}

The temperature-dependent Rabi frequency is then conveniently rewritten as
\begin{equation}
\begin{split}
    & \Omega_T = \Omega_0 \exp\left[ -F(T)  \right]
\end{split}
\end{equation}
with the new fit parameter, having dimensionality
\begin{equation}
    [\mathcal{J}\mathcal{E}^2] = 1/{\rm Hz}^b.
\end{equation}

\subsection{Formula wrap-up}
Interestingly, we can rewrite the shift and the dephasing rate in Eq.~\eqref{eq:shift_dephs_rate_final}, and the cross-section suppression function in Eq.~\eqref{eq:cross-section_suppression_F} in a very compact way, by defining the dimensionless scaling function 
\begin{equation}
    g(T, \omega, \nu) = \frac{1}{\zeta(\nu)}\sum_{n=1}^{\infty}\frac{1}{\left(n + k_BT/(\hbar \omega)\right)^{\nu}}.
\end{equation}
Here $\zeta(\nu)$ is the Riemann zeta function.
This function is limited to one, $g(T, \omega, \nu) \leq 1$ and goes to zero for $T\rightarrow \infty$ as $g(T, \omega, \nu)\sim 1/T^{\nu-1}$.

The three quantities are then rewritten as
\begin{equation}\label{eq:formula_wrapup}
    \begin{split}
        \omega_{\phi}(T) & = {\rm sign}(\mathcal{X})\,\frac{k_BT_{\omega}}{\hbar}\,\left(\frac{T}{T_{\omega}}\right)^{a+1}\, g(T, \omega_{\rm cut}, a+1),
        \\
        \gamma_{\phi}(T) & = \frac{k_BT_{\phi}}{\hbar}\left(\frac{T}{T_{\phi}}\right)^{2a+1}\, g\left(T, \frac{\omega_{\rm cut}}{2}, 2a+1\right),
        \\
        \Delta F(T) & = \left(\frac{T}{T_F}\right)^{b}\, g(T, \omega_{\rm cut}, b).
    \end{split}
\end{equation}
Here ${\rm sign}(\mathcal{X}) = \pm 1$ is the sign of the coefficient $\mathcal{X}$, which can give rise to a thermal blue/red shift.
The activation temperatures read
\begin{equation}
    \begin{split}
        k_B T_{\omega} & = \left(\frac{1}{\Gamma(a+1)\zeta(a+1)}\right)^{1/a}\frac{\hbar}{(\mathcal{J}|\mathcal{X}|)^{1/a}},
        \\
        k_B T_{\phi} & = \left(\frac{1}{\Gamma(2a+1)\zeta(2a+1)}\right)^{1/(2a)}\frac{\hbar \mathcal{K}^{1/(2a)}}{(\mathcal{J}\mathcal{X})^{2/(2a)}},
        \\
        k_BT_F & = \left(\frac{1}{\Gamma(b)\zeta(b)}\right)^{1/b}\frac{\hbar}{(\mathcal{J}\mathcal{E}^2)^{1/b}}.
    \end{split}
\end{equation}
The zero-point Huang-Rhys factor is 
\begin{equation}
    F_0 = \frac{1}{2\zeta(b)}\left(\frac{\hbar \omega_{\rm cut}}{k_BT_F}\right)^b,
\end{equation}
and represents the contribution to the Debye-Waller factor coming from the zero-point energy of the quantized phononic modes.


We remind that $a=s+p-1$ and $b=s+2r+1$, with $s\geq3$, $p\approx 1$, and $r\approx 1$.
From Eqs.~\eqref{eq:formula_wrapup}, we exctract the independent fit parameters
\begin{align}
    \mathcal{J}\mathcal{D} && (\mathcal{J}\mathcal{X})^2/\mathcal{K} && \mathcal{J}\mathcal{E}^2 && \omega_{\rm cut},
\end{align}
that can be equivalently replaced by the activation temperatures as fit parameters
\begin{align}
    T_{\omega} && T_{\phi} && T_{F} && \omega_{\rm cut}.
\end{align}

The exponents should follow approximately
\begin{align}
    a \geq 3 && b \geq 6,
\end{align}
but they can also be kept as free parameters and arranged to optimize the fit.

\section{Theory: a two-level molecule with dephasing and incoherent pump}
\label{sec:OBE}
As in the previous section, we treat the molecule as a perfect single-photon emitter, consisting of only the two HOMO-LUMO electronic levels $|g\rangle, |e\rangle$.
The molecule is driven by an external laser of frequency $\omega_L$ and with light intensity $I$. 
The resulting molecule's Rabi frequency is $\Omega_T(I)$, depending on the laser intensity $I$, and also depends on the environment temperature $T$ through the Debye-Waller factor, as described in Eq.~\eqref{eq:Rabi_DW}.

In the frame rotating with the laser frequency, the resonant drive Hamiltonian is
\begin{equation}\label{eq:zpl_drive_OBE}
    H_{\rm drive} = \hbar \Omega_T\left( \sigma + \sigma^{\dag}\right),
\end{equation}
with the molecule Hamiltonian as in Eq.~\eqref{eq:Hmol_0} (but discarding the vibrational degree of freedom for simplicity)
\begin{equation}
    H_{\rm mol} = \hbar \Delta_{\rm zpl} \sigma^{\dag}\sigma.
\end{equation}
Here we use $\sigma = |g\rangle \langle e |$ and $\Delta_{\rm zpl} = \omega_{\rm zpl} - \omega_L$ is the detuning between the ZPL frequency $\omega_{\rm zpl}$ (the HOMO-LUMO transition) and the laser frequency $\omega_L$.

The molecule is subjected to excitation loss $\gamma_{1}$ (also called inverse life-time) but also temperature-dependent dephasing $\gamma_{\phi}(T)$ (see Sec.~\ref{sec:pure_deph}).

The molecule can also be driven by an off-resonant laser, exciting the $e-g$ transition plus a vibrational quantum at the frequency $\omega_{\rm v}$.
This is realized by addressing one of the sideband peaks, visible in Fig.~\ref{fig:off-resonant saturation bis}, exciting the molecule with a laser tuned at the frequency $\omega_L \approx \omega_{\rm zpl}+\omega_{\rm v}$.
We call this driving scheme $0-1$ off-resonant. 
When the vibrational decay rate $\gamma_{\rm v}$ is the largest frequency scale in play, we can adiabatically eliminate this state, obtaining an effective incoherent pumping rate for the molecule~\cite{De_Bernardis2025}
\begin{equation}
    \Gamma_+ =  \frac{\eta^2\Omega_{T,{\rm off}}^2}{\gamma_{\rm v}} \frac{\gamma_{\rm v}^2}{4\Delta_{\rm off}^2 + \gamma_{\rm v}^2},
\end{equation}
where $\Delta_{\rm off} = \omega_{\rm zpl}+\omega_{\rm v} - \omega_L$ is the detuning between the laser frequency and the vibrational side-band peak, and $\eta$ is the so-called Franck-Condon factor~\cite{De_Bernardis2025}.
We also introduce the off-resonant pumping Rabi frequency $\Omega_{T,{\rm off}}$, to differentiate from the ZPL Rabi frequency appearing in Eq.~\eqref{eq:zpl_drive_OBE}, since it is linked to a physically different laser.
However, they have both the same temperature dependence, given by Eq.~\eqref{eq:Rabi_DW}.

Bringing all the terms together, the total master equation is
\begin{equation}\label{eq:master_eq}
    \partial_t \rho = -i/\hbar[H_{\rm mol}+H_{\rm drive}, \rho] + \gamma_{1}\mathcal{L}(\rho, \sigma) + \gamma_{\phi}\mathcal{L}(\rho, \sigma^{\dag}\sigma) + \Gamma_+\mathcal{L}(\rho, \sigma^{\dag})
\end{equation}
where $\mathcal{L}(\rho, c) =  c\rho c^{\dag} - \frac{1}{2}\lbrace{c^{\dag} c , \rho\rbrace}$.

\subsection{Optical Bloch equations}

Let's consider now each component of the $2\times 2$ density matrix, $\rho_{ee}, \rho_{eg}, \rho_{ge}$ (only three components, using $\rho_{gg} = 1 - \rho_{ee}$ from ${\rm Tr}[\rho] = 1$).
The master equation is reduced to the following system of equations
\begin{equation}
    \begin{split}
        \partial_t \rho_{ee} & = i \Omega_T \left( \rho_{eg} - \rho_{ge} \right) - \left(\gamma_{1} + \Gamma_+\right)\rho_{ee} + \Gamma_+
        \\
        \partial_t \rho_{eg} & = -i \Omega_T \left( 1 - 2\rho_{ee} \right) - i\Delta_{\rm zpl} \rho_{eg}  - \left(\frac{\gamma_{1}}{2} + \frac{\gamma_{\phi}}{2} + \frac{\Gamma_+}{2}\right)\rho_{eg}
        \\
        \partial_t \rho_{ge} & = i \Omega_T \left( 1 - 2\rho_{ee} \right) + i\Delta_{\rm zpl} \rho_{ge}  - \left(\frac{\gamma_{1}}{2} + \frac{\gamma_{\phi}}{2} + \frac{\Gamma_+}{2}\right)\rho_{ge}
    \end{split}
\end{equation}

Considering only the emission from the two-level transition, the photon flux rate is given by
\begin{equation}\label{eq:R_zpl}
    R = \gamma_{\rm rad}\rho_{ee},
\end{equation}
where $\gamma_{\rm rad}$ is the radiative decay rate, for which holds $\gamma_{\rm rad}\approx \gamma_1$ for Fourier-limited molecules~\cite{toninelliSingleOrganicMolecules2021}.
In CW driving conditions, we can explicitly solve the system of equations, looking at the steady state by setting $\partial_t \rho = 0$ and solving the algebraic system
\begin{equation}\label{eq:OBE_ss}
    \begin{pmatrix}
        \gamma_{1} && -i\Omega_T && i\Omega_T\\
        -2i\Omega_T && i\Delta_{\rm zpl} + \gamma_{2} && 0\\
        2i\Omega_T && 0 && -i\Delta_{\rm zpl} + \gamma_{2}\\
    \end{pmatrix}
    \begin{pmatrix}
        \rho_{ee}\\
        \rho_{eg}\\
        \rho_{ge}\\
    \end{pmatrix}
    = 
    \begin{pmatrix}
        \Gamma_+\\
        -i\Omega_T\\
        i\Omega_T\\
    \end{pmatrix}
\end{equation}
Here, the total coherence decay rate is
\begin{equation}
    \gamma_{2}(T) = \frac{\gamma_{1} + \Gamma_{+} + \gamma_{\phi}(T)}{2},
\end{equation}
that explicitly depends on temperature through the dephasing rate.


We immediately notice that the incoherent pumping rate $\Gamma_+$ is a direct source for the excited population $\rho_{ee}$, while the Rabi frequency $\Omega_T$ due to the coherent drive (zpl resonant pump) is the source of the coherences $\rho_{eg},\rho_{ge}$ (i.e. the off-diagonal element of the density matrix), and the pump of $\rho_{ee}$ is only through the coupling with the coherences.
Moreover, the dephasing rate $\gamma_{\phi}(T)$ influences only the decay of coherences $\gamma_{2}(T)\sim \gamma_{\phi}(T)/2$, and it does not enter into the population decay rate $\gamma_{1}$.

\subsection{Off-resonant pumping}
\label{sec:OBE_off_res}
We consider here the $0-1$ off-resonant pumping scheme, which is effectively described by the incoherent pump term controlled by $\Gamma_+>0$, with zero coherent drive $\Omega_T=0$ in Eqs.~\eqref{eq:OBE_ss}.
The molecule's excited state population is then given
\begin{equation}\label{eq:rhoee_off}
    \rho_{ee}^{\rm off} = \frac{S^{\rm off}}{1+S^{\rm off}},
\end{equation}
where the incoherent (off-resonant) saturation parameter is
\begin{equation}
    S^{\rm off} = \frac{\Gamma_+}{\gamma_{1}}.
\end{equation}
The coherence terms $\rho_{eg}, \rho_{ge}$ are completely decoupled from the population and decay to zero. 
The population is thus not affected by any temperature dependence affecting the coherence decay rate $\gamma_2(T)$.
According to the molecule thermal theory described in Sec.~\ref{sec:thermal_theory}, the incoherent saturation parameter can still depend on the temperature $T$ only through the Debye-Waller factor, re-expressing the incoherent drive rate in terms of the underlying off-resonant Rabi frequency
\begin{equation}
   \frac{\Gamma_+}{\gamma_1} = \frac{I}{I_{\rm sat}^{\rm off}} e^{-2F(T)},
\end{equation}
where, using Eq.~\eqref{eq:Rabi_bare},
\begin{equation}\label{eq:Isat_off}
    I_{\rm sat}^{\rm off} = \frac{\gamma_1\gamma_{\rm v} e^2\hbar}{8\pi\alpha_{\rm fs} \eta_{\rm v}^2d_{\rm DBT}^2} \frac{4\Delta_{\rm off}^2 + \gamma_{\rm v}^2}{\gamma_{\rm v}^2}. 
\end{equation}

Following the measurement procedure described in the main text, the photon flux is measured through the emission via the ZPL, as in Eq.~\eqref{eq:R_zpl}, giving the photon flux rate~\cite{De_Bernardis2025}
\begin{equation}\label{eq:R_offRes}
    R^{\rm off} = \gamma_{\rm rad} \frac{S^{\rm off}}{1+S^{\rm off}}.
\end{equation}

\subsection{Resonant pumping}
\label{sec:OBE_res_pump}
For resonant pumping only, we take $\Gamma_+ = 0$, and $\Omega_T>0$. The system of equations in Eqs.~\eqref{eq:OBE_ss} can be easily solved, obtaining
\begin{equation}\label{eq:rhoee_res}
    \rho_{ee}^{\rm res} = \frac{1}{2}\frac{S^{\rm res}}{\Delta_{\rm zpl}^2/(4\gamma_2^2)+ 1 + S^{\rm res}}.
\end{equation}
Here, the resonant saturation parameter is (making explicit the intensity and temperature dependencies)
\begin{equation}\label{eq:Sres}
    S^{\rm res}(I, T) = \frac{2\Omega_T^2(I)}{\gamma_{2}(T)\gamma_{1}} = \frac{I}{I_{\rm sat}^{\rm res}}\frac{\gamma_1}{2\gamma_2(T)}e^{-2F(T)}.
\end{equation}
The saturation intensity is defined using Eq.~\eqref{eq:Rabi_bare} as
\begin{equation}\label{eq:I_sat_res}
    I_{\rm sat}^{\rm res} = \frac{\gamma_1^2 e^2\hbar}{2\pi\alpha_{\rm fs}d_{\rm DBT}^2}.
\end{equation}
The linewidth of the Lorentzian population in Eq.~\eqref{eq:rhoee_res}, scanning over $\Delta_{\rm zpl}/(2\pi)$ (the frequency, not the pulsation), is given by
\begin{equation}\label{eq:linewidth_zpl}
    \Gamma_{\rm zpl}(I, T) =  \frac{\gamma_2(T)}{\pi}\sqrt{1 + S^{\rm res}( I, T)},
\end{equation}
exhibiting the typical power broadening dependence on the saturation parameter.

From Eq.~\eqref{eq:Sres} it is thus clear that the resonant saturation parameter is inversely proportional to the dephasing rate $S^{\rm res} \sim 1/\gamma_{\phi}$.
In this way, the fluorescence can be suppressed both by increasing the dephasing rate and also (as in the incoherent pump case) by decreasing the Debye-Waller factor.

Since, under the resonant pumping scheme, the photon flux is typically detected through the red-shifted fluorescence (see the red-sideband peaks, for instance, in Fig.~\ref{fig:off-resonant saturation bis}), the resulting rate is multiplied by a Franck-Condon factor
\begin{equation}
    R^{\rm res} \approx \gamma_{\rm rad} \eta_{\rm tot}^2 \rho_{ee},
\end{equation}
which is in general the quadratic sum $\eta_{\rm tot}^2 = \sum_{n}\eta_n^2$ of the individual Franck-Condon factor of each $n$-th vibrational mode in the detected bandwidth~\cite{De_Bernardis2025}.
The quantity $\eta_{\rm tot}^2$ is related to the fraction of photons emitted in the ZPL, via $1-\eta_{\rm tot}^2$~\cite{toninelliSingleOrganicMolecules2021}, and for DBT molecules is around $\eta_{\rm tot}^2\sim 0.1-0.2$. Compared to the off-resonant photon flux in Eq.~\eqref{eq:R_offRes}, the measured photon flux under resonant drive is thus one order of magnitude smaller.

\subsection{Extracting the Debye-Waller factor}

While the increase of dephasing with temperature $\gamma_{\phi}(T)$ can fully explain the resonant anomalous saturation, the off-resonant (incoherent) pump requires a strong temperature-dependent Debye-Waller factor $e^{-F(T)}$.
It is thus crucial to define how to extract the Debye-Waller (DW) factor from the experimental data under coherent or incoherent pump.

In our experiment, we focused mainly on resonant pumping, and so we do here with the theory.
Combining the expressions in Eqs.~\eqref{eq:rhoee_res}-\eqref{eq:Sres}-\eqref{eq:linewidth_zpl}, we find
\begin{equation}
    e^{-2F(T)} = \frac{S^{\rm res}(I, T)}{\sqrt{1+S^{\rm res}(I, T)}} \frac{2\pi\Gamma_{\rm zpl}(I, T) }{\gamma_1}\frac{I_{\rm sat}^{\rm res}}{I}.
\end{equation}
All the quantities appearing in this expression can be extracted from the crossing of different experimental data.
$\gamma_1$ and $I_{\rm sat}^{\rm res}$ can be obtained by low temperature, low intensity scans of the ZPL.
The saturation parameter is extracted by considering its relation with the measured photon flux rate $R_{\rm exp}$ at full resonance $\Delta_{\rm zpl}=0$
\begin{equation}
    S^{\rm res}(I, T) = \frac{R_{\rm exp}/R_{\infty}}{1-R_{\rm exp}/R_{\infty}}.
\end{equation}
Here $R_{\infty} = R_{\rm exp}(\Delta_{\rm zpl} = 0, I\rightarrow\infty, T=0)$ is the photon flux rate obtained in resonant pump, at full saturation, at zero temperature.

\section{Theory: a dark-state model for excited state absorption}
\label{sec:theory_ESA}
Leaving the hypothesis that the anomalous saturation is generated by thermal heating, we introduce a model based on the presence of a highly dissipative dark state.
Let's assume that our previous model was incomplete and must be supplemented with an additional energy level for the molecule. In this way, we deal with a three-level system, where the third state is called $|d\rangle$, i.e., a dark state. 
We assume that the system can access the dark state only through the excited electronic state, and so we call this process \emph{excited state absorption} (ESA).

\subsection{Resonant pumping}
\label{sec:res_pump_dark}
The system Hamiltonian is rewritten as 
\begin{equation}
    H_0 = \hbar\Delta_{\rm zpl}|e\rangle \langle e | + \hbar\Delta_{d}|d\rangle \langle d |,
\end{equation}
\begin{equation}\label{eq:drive_dark_res}
    H_D = \hbar\Omega_T \left( |g\rangle \langle e | + |e\rangle \langle g | \right) + \hbar\Omega_{d} \left( |e\rangle \langle d | + |d\rangle \langle e | \right).
\end{equation}

Assuming that the $|d\rangle$ state decays very quickly with rate $\gamma_d\gg 2\gamma_2, \gamma_{\rm v}$ to the ground state $|g\rangle$, we adiabatically eliminate it (with the standard procedure), obtaining an effective Lindbladian dissipator for the master equation in Eq.~\eqref{eq:master_eq}
\begin{equation}
    \gamma_{\rm ESA}\mathcal{L}(\rho, |g\rangle\langle e |).
\end{equation}
Here, the effective decay rate generated by the ESA is
\begin{equation}
    \gamma_{\rm ESA}(I) = \frac{\Omega_d^2(I)}{\gamma_d}\frac{\gamma_d^2}{4\Delta_d^2+\gamma_d^2}.
\end{equation}
Assuming a single-photon transition we can rewrite this expression as
\begin{equation}\label{eq:ESA_gamma_SI}
     \gamma_{\rm ESA}(I) = \gamma_1 \frac{I}{I_{\rm ESA}},
\end{equation}
where the ESA saturation intensity sets the threshold intensity of the anomalous saturation, and, assuming the dark state Rabi frequency with the same shape as Eq. \eqref{eq:Rabi_bare}, reads
\begin{equation}
    I_{\rm ESA} =  e^2\hbar\frac{\gamma_1\gamma_d}{8\pi\alpha_{\rm fs} d_{\rm dark}}\frac{4\Delta_d^2+\gamma_d^2}{\gamma_d^2},
\end{equation}
where $d_{\rm dark}= \langle {e| \hat{d}|{d}} \rangle$ is the matrix element from the molecule excited state to the dark state (replacing $d_{\rm DBT}$ in Eq. \eqref{eq:Rabi_bare}).

In this way, the decrease in the saturation parameter in Eq.~\eqref{eq:Sres} is mostly given by an increase in the population loss rate, given by 
\begin{equation}
    \tilde{\gamma}_1 = \gamma_{1} + \gamma_{\rm ESA}
\end{equation}
At resonance, and considering the dephasing rate $\gamma_{\phi}\approx 0$ (at very small temperature), the saturation parameter is modified as
\begin{equation}
    \tilde{S}^{{\rm res}} = \frac{S^{{\rm res}}}{\left(1+\gamma_{\rm ESA}(I)/\gamma_{1}\right)^2},
\end{equation}
and thus the saturation curve can be rewritten as
\begin{equation}
    \tilde{R}^{\rm res} = R^{\rm res}_{\infty}\frac{S^{\rm res}}{(1+\xi^{\rm res} S^{\rm res})^2 + S^{\rm res}}.
\end{equation}
The anomaly parameter is then defined as
\begin{equation}
    \xi^{\rm res} = \frac{\gamma_{\rm ESA}(I^{\rm res}_{\rm sat})}{\gamma_{1}}.
\end{equation}

\subsection{Incoherent pumping}
In the presence of off-resonant pump, we must take into consideration that the system passes through a dissipative vibrational level, but we assume that the dark state $d$ can be reached in the same way.  
The transition triggered by the off-resonant laser is $|g,0_{\rm v}\rangle \mapsto |e, 1_{\rm v}\rangle$, which is at the frequency $\sim \omega_0 + \omega_{\rm v}$.
To be resonant, as in the resonant pumping case, we assume that the transition to the dark state also involves a vibrational level $|e, 1_{\rm v}\rangle \mapsto |d, 2_{\rm v}\rangle$.
Then the dark state is highly dissipative and decays to the ground state $|d, 1_{\rm v}\rangle \mapsto |g,0_{\rm v}\rangle$.
The Hamiltonian describing the overall drive is
\begin{equation}
    H_{D'} = \hbar\eta_{\rm v}\Omega_T\left( |g,0_{\rm v}\rangle \langle e, 1_{\rm v}| + |e,1_{\rm v}\rangle \langle g, 0_{\rm v}| \right) + \hbar\Omega_{d'}\left( |e, 1_{\rm v}\rangle \langle d, 2_{\rm v}| + |d, 2_{\rm v}\rangle \langle e,1_{\rm v}| \right).
\end{equation}
This driving Hamiltonian is different from the one employed in the resonant description in Eq. \eqref{eq:drive_dark_res} for two reasons.
The first term has amplitude multiplied by the Franck-Condon factor $\eta_{\rm v}$, indicating that a vibrational transition is also involved.
The second term has Rabi frequency $\Omega_{d'}\neq \Omega_d$, in principle, also due to involving a vibrational transition too.
The ratio $\Omega_{d'}/\Omega_d = \langle d, 2_{\rm v}| \hat{d} |e, 1_{\rm v}\rangle / \langle d, 0_{\rm v}| \hat{d} |e, 0_{\rm v}\rangle = \sqrt{2}\eta_d$ is essentially the Franck-Condon factor associated to this dark-state transition.
Here, for simplicity, we assume it $\eta_d \approx 1/\sqrt{2}$, such that $\Omega_{d'}\approx\Omega_d$.
In this way we can proceed with the same approach as in Sec. \ref{sec:res_pump_dark}, using the same $\gamma_{\rm ESA}$.

We adiabatically eliminate the dark state, as in the resonant pumping case, ending up with a three-level system, composed of $|g,0_{\rm v}\rangle, |e,0_{\rm v}\rangle, |e,1_{\rm v}\rangle$. Assuming, for simplicity, zero detunings, the Hamiltonian is
\begin{equation}
    H_{\rm 3LS} = \hbar\eta_{\rm v}\Omega_T\left( |g,0_{\rm v}\rangle \langle e, 1_{\rm v}| + |e,1_{\rm v}\rangle \langle g, 0_{\rm v}| \right).
\end{equation}
We also have three main sources of dissipation, given by the jump operators
\begin{align}
    \hat{L}_- = \sqrt{\gamma_{\rm ESA}}|g,0_{\rm v}\rangle \langle e, 1_{\rm v}| && \hat{L}_{\rm v} = \sqrt{\gamma_{\rm v}}|e,0_{\rm v}\rangle \langle e, 1_{\rm v}| && \hat{L}_1 = \sqrt{\gamma_{1}}|g,0_{\rm v}\rangle \langle e, 0_{\rm v}|.
\end{align}
We adopt the convention that each of these jump operators $\hat{L}$ enters in the master equation as $\partial_t \hat{\rho} = \ldots + \hat{L}\hat{\rho}\hat{L}^{\dag} -\lbrace{\hat{L}^{\dag}\hat{L}, \hat{\rho}\rbrace}$.
Due to the existence of the dark state, the excited-vibrational state $|e,1_{\rm v}\rangle$ now has two distinct decay channels.
We can thus use the method in  Ref.~\cite{soresen_PhysRevA.85.032111} to further eliminate this state, and remain with an effective two-level description.
The system has purely dissipative dynamics, given now by the jump operators
\begin{align}\label{eq:jump_effTLS}
    \hat{L}_1 = \sqrt{\gamma_{1}}|g,0_{\rm v}\rangle \langle e, 0_{\rm v}| && \hat{L}_+ = \sqrt{\tilde{\Gamma}_{+}}|e,0_{\rm v}\rangle \langle g, 0_{\rm v}|.
\end{align}
Here we introduced the ESA-modified incoherent pump
\begin{equation}
    \tilde{\Gamma}_{+} = \frac{\eta_{\rm v}^2\Omega_T^2}{\gamma_{\rm v}}\frac{\gamma_{\rm v}^2}{(\gamma_{\rm v}+\gamma_{\rm ESA}(I))^2}.
\end{equation}
Here the ESA loss rate is the same as in Eq. \eqref{eq:ESA_gamma_SI} considering $\Delta_{d} = 0$.
Notice that the first contribution on the right-hand-side is the standard incoherent pumping rate~\cite{De_Bernardis2025}, while the second contribution is a branching ratio quantifying how much excitation goes effectively into the excited state instead of being absorbed by the dark state.

The steady state of the master equation defined by the jump operators in Eq. \eqref{eq:jump_effTLS} can be easily solved for its ground state, finding
\begin{equation}
    \rho_{ee}^{\rm off} = \frac{S^{\rm off}}{(1+\xi^{\rm off}S^{\rm off})^2 + S^{\rm off}},
\end{equation}
where
\begin{equation}
    S^{\rm off} = \frac{\eta_{\rm v}^2\Omega_T^2}{\gamma_1\gamma_{\rm v}},
\end{equation}
and
\begin{equation}
    \xi^{\rm off} = \frac{\gamma_1}{\gamma_{\rm v}}\frac{I^{\rm off}_{\rm sat}}{I_{\rm ESA}}.
\end{equation}

\newpage
\section{Reduction of inter-system crossing rate at higher excitation intensities}
\label{sec:ISC_reduction}

\begin{figure}
    \centering
    \begin{subcaptiongroup}
        \subcaptionlistentry{triplet linewidth}
        \label{fig:sup:triplet:linewidth}
        \subcaptionlistentry{triplet g2 short time low power}
        \label{fig:sup:triplet:short:low}
        \subcaptionlistentry{triplet g2 short time high power}
        \label{fig:sup:triplet:short:high}
        \subcaptionlistentry{triplet g2 long time}
        \label{fig:sup:triplet:long}
        \subcaptionlistentry{triplet isc}
        \label{fig:sup:triplet:isc}
    \end{subcaptiongroup}
    \includegraphics[width=\linewidth]{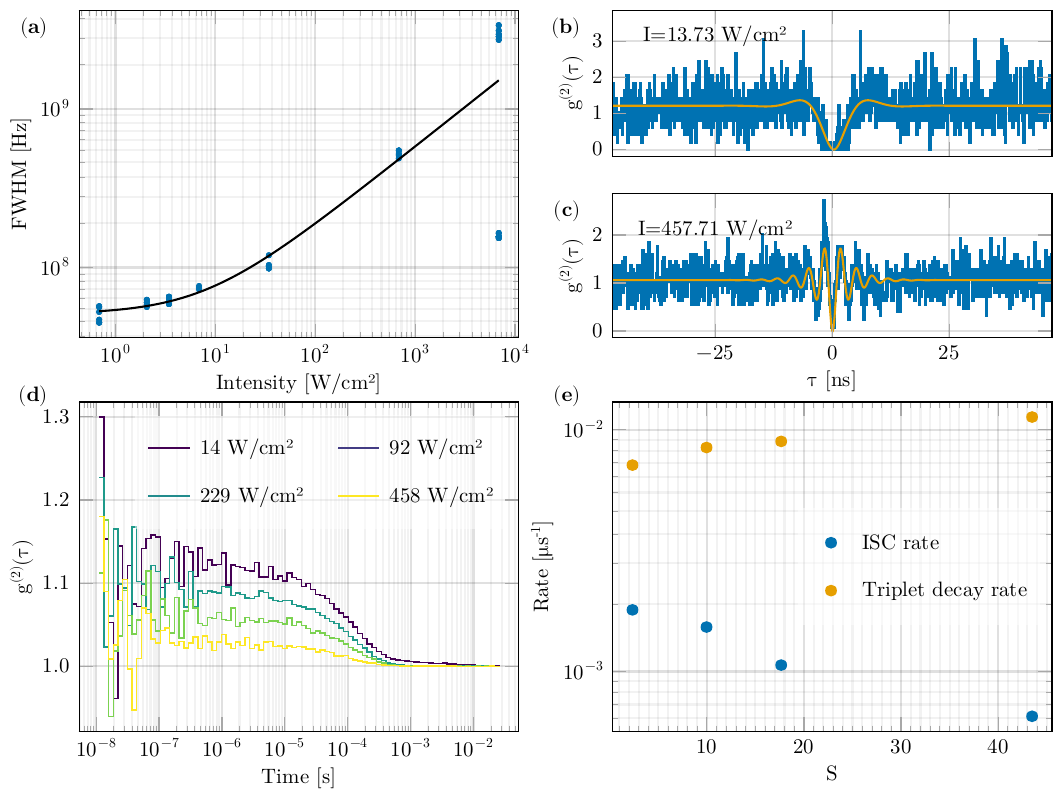}
    \caption{\subref{fig:sup:triplet:linewidth}~Linewidth of the ZPL transition of a molecule as a function of excitation laser intensity (blue dots), fitted with the expected law for a two-levels system (black line). \subref{fig:sup:triplet:short:low}~and~\subref{fig:sup:triplet:short:high}~Second-order correlation function ($g^{(2)}(\tau)$) at short time-scales of the red-shifted fluorescence of the same molecule when excited resonantly at various laser intensities. \subref{fig:sup:triplet:long}~Second-order correlation function at long time-scales of the red-shifted fluorescence of the molecule for varying excitation intensities. \subref{fig:sup:triplet:isc}~Characteristic inter-system crossing rates and triplet decay rates extracted from \subref{fig:sup:triplet:long}.}
    \label{fig:sup:triplet}
\end{figure}

An accurate mapping of inter-system crossing rate, \textit{i.e.} of the transition rate from the excited singlet state towards a triplet state, can be realized using photon statistics. In this experiment, we first realized a measurement of the saturation and broadening of the ZPL, as in the main text, to extract the decoherence rate $\gamma_2=1/T_2=1/2T_1 + 1/T_\phi$ ($T_\phi$ the pure dephasing time) of the system, as shown in Fig.~\ref{fig:sup:triplet:linewidth}. Then, we measured the second-order correlation function (denoted $g^{(2)}(\tau)$) of the Stokes-shifted fluorescence both at short time scales (Fig.~\ref{fig:sup:triplet:short:low} and~\ref{fig:sup:triplet:short:high}), and at long time scales (Fig.~\ref{fig:sup:triplet:long}). 

First, the short-time $g^{(2)}(\tau)$ measurements can be fitted with the general formula given in Ref.~\cite{Grandi2016}:
\begin{align}
g^{(2)}(\tau) &= 1 - \frac{p+q}{2q}e^{-\frac{1}{2}(p-q)\tau} + \frac{p-q}{2q}e^{-\frac{1}{2}(p+q)\tau}\\
p &= \gamma_1 + \gamma_2\\
q &= \sqrt{(\gamma_1-\gamma_2)^2-4\Omega^2}
\end{align}
Where $\gamma_1=1/T_1$ is the decay rate of the excited state ($\sim\SI{4}{ns}$), $\gamma_2$ has been defined in the previous paragraph, and $\Omega$ is the Rabi frequency. It is convenient to rewrite $\Omega$, using the saturation parameter $S$ as a proxy, with $S=\Omega^2/(\gamma_1\gamma_2)$. This means we can rewrite $q$ as
\begin{equation}
q = \sqrt{(\gamma_1-\gamma_2)^2-4S\gamma_1\gamma_2}
\end{equation}
Leaving only the $S$ parameter free in the fit procedure yields a setup-independent measurement of the saturation parameter. Two examples of the result of this fit procedure is given in Fig.~\ref{fig:sup:triplet:short:low} and~\ref{fig:sup:triplet:short:high}.

Second, the behavior at long time of the $g^{(2)}(\tau)$ function can be fitted using the formula given in Ref.~\cite{Bernard1993a}. In particular, this formula gives a way to account for the effect of the background on the $g^{(2)}(\tau)$ measurement, by expressing the measured intensity correlation as 
\begin{equation}
\langle I(t)I(t+\tau) \rangle = b^2 + 2bs + \langle s(t)s(t+\tau) \rangle = b^2 + 2bs + s^2g^{(2)}(\tau)
\end{equation}
where $g^{(2)}(\tau)$ is the second-order correlation function of the fluorescence from the molecule, $s$ is the emission rate of the molecule, and $b$ is the detection rate due to the background. The three terms of the addition correspond to a detection of two background photons, a background and a signal photons, and of two signal photons. Fortunately, $b$ and $s$ are easily extracted from a fit of the shape of the ZPL, $b$ corresponding to the baseline and $s$ to the amplitude of the line. The corrected experimental $g^{(2)}(\tau)$ functions are shown in Fig.~\ref{fig:sup:triplet:long} for various excitation laser intensities. At long time scales (typically $\tau>10^{-7}\;\text{s}$ in our case), Bernard and coworkers~\cite{Bernard1993a} showed that the second-order correlation function can be re-written as in Equation~\ref{eq:g2 long time},
\begin{equation}
g^{(2)}(\tau) = 1 + ae^{-\lambda\tau} \label{eq:g2 long time}
\end{equation}
with
\begin{align}
\lambda &= \gamma_\text{T} \left(1 + a\right)\\
a &= \frac{S}{1+2S\gamma_\text{T}/\gamma_{\text{isc}}}
\end{align}
where $\gamma_\text{T}$ is the triplet de-excitation rate, and $\gamma_\text{isc}$ is the intersystem crossing rate.

Fitting the $g^{(2)}(\tau)$ functions at long time in Fig.~\ref{fig:sup:triplet:long} with Equation~\ref{eq:g2 long time} leaving only $\lambda$ and $a$ free ($S$ being given by the fit of the $g^{(2)}(\tau)$ function at short time) allows the extraction of the inter-system crossing rate $\gamma_\text{isc}$ and triplet de-excitation rate $\gamma_\text{T}$ as a function of the saturation parameter $S$. For convenience, we show the corresponding characteristic times in Fig.~\ref{fig:sup:triplet:isc}. It is clear that the inter-system crossing time (blue dots) is increasing (from \SI{500}{\micro\second} to \SI{1500}{\micro\second})  a function of the saturation parameter $S$, while the triplet lifetime is more stable in comparison (decreasing from \SI{140}{\micro\second} to \SI{88}{\micro\second}). 

We interpret this increase of the inter-system crossing time as a side-effect of excited state absorption. In this picture, the excited-state absorption competes with the transition to the triplet, making the inter-system crossing less and less likely, which results in a longer apparent inter-system crossing time.

\newpage
\section{Correlation with optical shift}
\label{sec:optical_shift}

\begin{figure}
    \centering
    \includegraphics{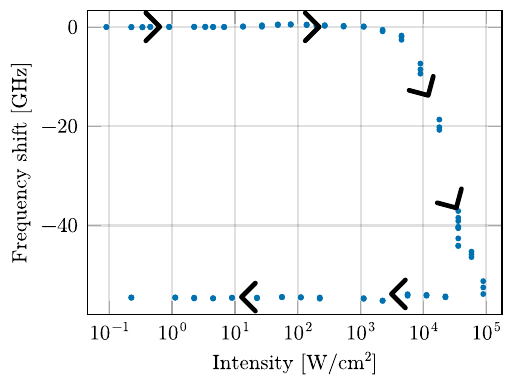}
    \caption{Central frequency of the ZPL as a function of laser excitation intensity (blue dots). The black arrows indicate the order in which the successive measurements were performed.}
    \label{fig:sup:shift}
\end{figure}
As the experiments reported in this work involve shining laser light at high intensity on the samples, we observe optical Stark shift~\cite{colauttiLaserInducedFrequencyTuning2020, Duquennoy2024} in conjunction to the reported anomalous saturation. Supl.~Fig.~\ref{fig:sup:shift} shows the evolution of the position of the central frequency of the ZPL as the excitation intensity is varied up and down. The black arrows illustrate the time evolution. It is clear that there is a hysteresis ($\sim\SI{54}{GHz}$) of the frequency shift as power is being varied. We argue that this simultaneity of the two experimental results is, actually, more than a mere coincidence. 

\begin{figure}
    \centering
    \begin{subcaptiongroup}
        \subcaptionlistentry{Sublimated saturation amplitude}
        \label{fig:sup:sublimated:saturation amplitude}
        \subcaptionlistentry{Sublimated saturation linewidth}
        \label{fig:sup:sublimated:saturation linewidth}
        \subcaptionlistentry{Sublimated saturation shift}
        \label{fig:sup:sublimated:saturation shift}
    \end{subcaptiongroup}
    \includegraphics[width=\linewidth]{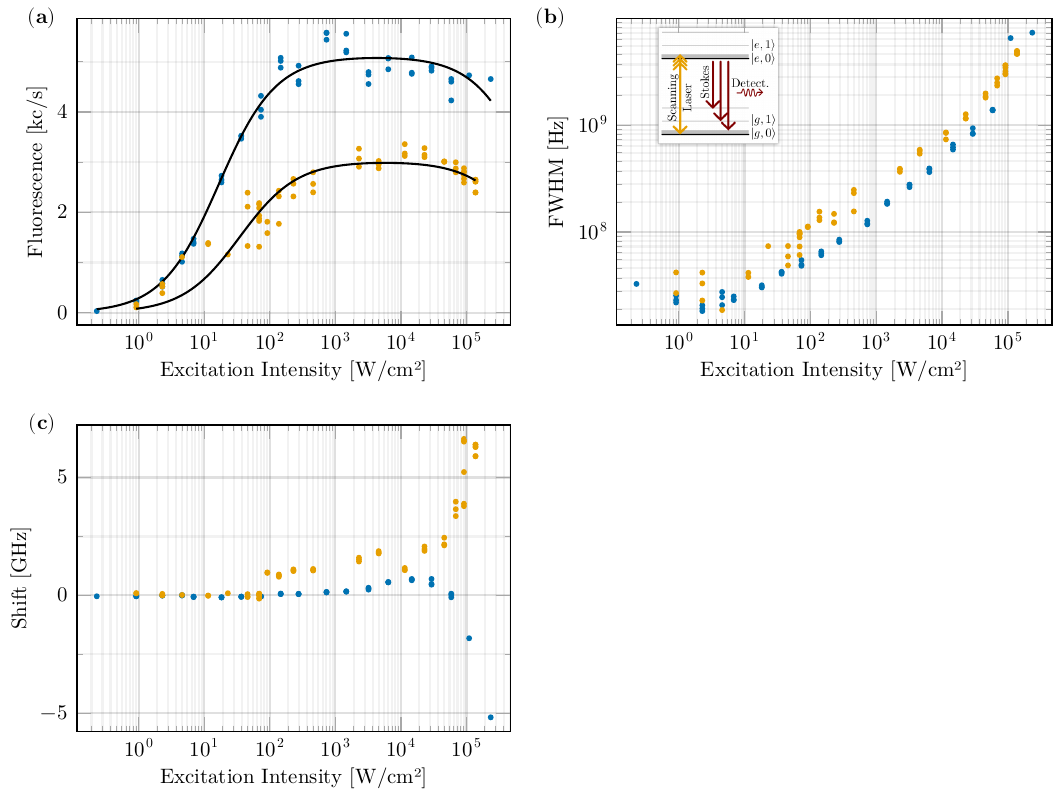}
 34    \caption{\subref{fig:sup:sublimated:saturation amplitude}~Amplitude of the detected resonance peak of two molecules in sublimated crystals as the intensity of the excitation laser is being scanned (dots). Each dataset has been fitted with the anomalous saturation model given in Equation~\ref{eq:anomalous_saturation} (solid black lines). The resulting fit parameters are $I_{\text{sat}}^{\text{res}}=\SI{16.7}{W/cm^2}$, $R_{\infty}^{\text{res}}=\SI{5150}{kc/s}$, $\xi^{\text{res}}=0.0039$, and $I_{\text{sat}}^{\text{res}}=\SI{35.3}{W/cm^2}$, $R_{\infty}^{\text{res}}=\SI{3060.0}{kc/s}$, $\xi^{\text{res}}=0.0062$. \subref{fig:sup:sublimated:saturation linewidth}~Corresponding linewidth of the transition. The inset shows the excitation scheme. \subref{fig:sup:sublimated:saturation shift}~Corresponding shift of the position of the ZPL. }
    \label{fig:sup:sublimated}
\end{figure}

We believe that the anomalous saturation regime is enabled by the matrix itself, and is due in particular to the quality of the embedding within the crystal. To test this theory, we moved to sublimated crystals of anthracene. These much bigger crystals, typically hundreds of micrometers wide, provide a much better crystal quality than the anthracene nanocrystals obtained by reprecipitation used in the main text. This is typically illustrated by the very narrow inhomogeneous broadenings that are reported in the literature~\cite{Lange2026}. In Suppl.~Fig.~\ref{fig:sup:sublimated}, we report intensity scan experiments performed on sublimated Ac crystals. The first dataset (blue dots) shows no clear sign of the anomalous saturation regime within the range of scanned intensities, while the onset of it is visible in the second dataset (orange dots). This second dataset was purposely taken on the edge of a crystal in order to find a molecule that would present an embedding of lesser quality. We also note that, in both cases, the optical Stark shift is rather small (Supl.~Fig.~\ref{fig:sup:sublimated:saturation shift}), typically $\sim$\SI{5}{GHz}, compared to the shift observed in nanocrystals (typically \SI{50}{GHz}). This illustrates the consequences of the environment's quality on the magnitude of the anomalous saturation regime.

\clearpage
\newpage

\bibliography{bibliography}

\end{document}